%% file: main_.tex
\documentclass{article}
\PassOptionsToPackage{numbers}{natbib}
\usepackage[preprint]{nips_2018}
\usepackage{authblk}

\usepackage[log-declarations=false]{xparse}
\usepackage[final]{physicist-taper}
\usepackage{qcircuit}
\usepackage{algpseudocode}
\usepackage[title]{appendix}
\usepackage[caption=false]{subfig}

\begin{document}

\title{Universal discriminative quantum neural networks
}

\date{\today}
\newcommand{\UCLPhy}{Department of Physics \& Astronomy, University College London, London, UK}
\newcommand{\UCLCS}{Department of Computer Science, University College London, London, UK}
\newcommand{\SJTU}{Institute of Natural Sciences, Shanghai Jiao Tong University, Shanghai, China}
\newcommand{\GoogleQAI}{Google Quantum AI Laboratory, Venice, California, USA}
\newcommand{\Achenhx}{H. Chen}
\newcommand{\ALeo}{L. Wossnig}
\newcommand{\ALeoEmail}{l.wossnig@cs.ucl.ac.uk}
\newcommand{\ASimone}{S. Severini}
\newcommand{\AHartmut}{H. Neven}
\newcommand{\AMasoud}{M. Mohseni}
\newcommand{\AMasoudEmail}{mohseni@google.com}


\renewcommand\Authfont{\textbf}
\author[1,2]{\Achenhx}
\author[2,*]{\ALeo}
\author[2,3]{\ASimone}
\author[4]{\AHartmut}
\author[4,$\dagger$]{\AMasoud} 
\affil[1]{\UCLPhy}
\affil[2]{\UCLCS} 
\affil[3]{\SJTU}
\affil[4]{\GoogleQAI} 
\let\oldthefootnote\thefootnote
\renewcommand{\thefootnote}{\fnsymbol{footnote}}
\footnotetext[1]{Email: \url{\ALeoEmail}}
\footnotetext[2]{Email: \url{\AMasoudEmail}}
\let\thefootnote\oldthefootnote

\maketitle

\begin{abstract}
Quantum mechanics fundamentally forbids deterministic discrimination of quantum states and processes. However, the ability to optimally distinguish various classes of quantum data is an important primitive in quantum information science. In this work, we train near-term quantum circuits to classify data represented by non-orthogonal quantum probability distributions using the Adam stochastic optimization algorithm. This is achieved by iterative interactions of a classical device with a quantum processor to discover the parameters of an unknown non-unitary quantum circuit. This circuit learns to simulates the unknown structure of a generalized quantum measurement, or Positive-Operator-Value-Measure (POVM),  that is required to optimally distinguish possible distributions of quantum inputs. Notably we use universal circuit topologies, with a theoretically motivated circuit design, which guarantees that our circuits can in principle learn to perform arbitrary input-output mappings. Our numerical simulations show that shallow quantum circuits could be trained to discriminate among various pure and mixed quantum states exhibiting a trade-off between minimizing erroneous and inconclusive outcomes with comparable performance to theoretically optimal POVMs. We train the circuit on different classes of quantum data and evaluate the generalization error on unseen mixed quantum states. This generalization power hence distinguishes our work from standard circuit optimization and provides an example of quantum machine learning for a task that has inherently no classical analogue.
\end{abstract}

\input{sections/intro-qsd.tex}
\input{sections/results.tex}

\newpage
\appendix
\input{sections/appendix.tex}

\bibliographystyle{ieeetr}
\bibliography{library.bib}{}

\end{document}

%% file: sections/intro-qsd.tex
\section{Introduction}

The interface of quantum physics and machine learning has recently received a considerable amount of interest. Two complementary  methodologies have been developed to address the question whether quantum mechanics can help with solving machine learning tasks or similarly whether machine learning techniques could help solving problems in quantum computation and many-body condensed matter systems more efficiently ~\cite{ciliberto2018quantum,biamonte2017quantum,schuld2015introduction}.

The circuit model of universal fault-tolerant quantum computers has been shown to offer a range of machine learning algorithms~\cite{lloyd2014pca,lloyd2013supervised,rebentrost2014quantum,wiebe2015quantum} which could lead to considerable speed-ups under certain assumptions over classical counterparts.
The underlying property which results in such quantum advantage is the ability of quantum computers to perform certain linear algebra operations faster than classical machines~\cite{harrow2009quantum,wossnig2017quantum,childs2017quantum,ambainis2010variable,dervovic2018quantum}. The algorithmic complexity for some of these quantum learning schemes is in principle $O(\text{polylog}(N))$ for input dimension $N$ instead of the classical $O(\text{poly}(N))$ which can manifest itself as exponential speedup when applied to quantum data~\cite{lloyd2014pca}. However, these algorithms have been shown to come with certain caveats, when applied to classical data. The most relevant is the preparation of quantum states that encode classical information and is believed to have in the worst case a polynomial scaling in the input dimension of the data~\cite{aaronson2015read,childs2009quantum} diminishing their advantages in the first place.\\
Here we take a different approach and utilize a hybrid quantum-classical setup to directly learn the design of shallow quantum circuits for an inherently quantum-mechanical task, which has no classical counterpart based on the non-orthogonal nature of quantum states. Hence we solve a problem where a comparison with classical algorithms is not appropriate unlike for the case of the above mentioned algorithms.
Recent works focused on using quantum-classical hybrid methods for training quantum circuits for a range of tasks~\cite{wan2017quantum,romero2017quantum,mitarai2018quantum,farhi2018classification,verdon2017quantum,li2017efficient,grant2018hierarchical,schuld2018circuit}. However, prior \textit{quantum circuit training} works rarely give a motivation of the underlying circuit structure nor provide a concrete quantum application. 

\begin{figure}[ht]
\centering
\includegraphics[width=0.6\textwidth]{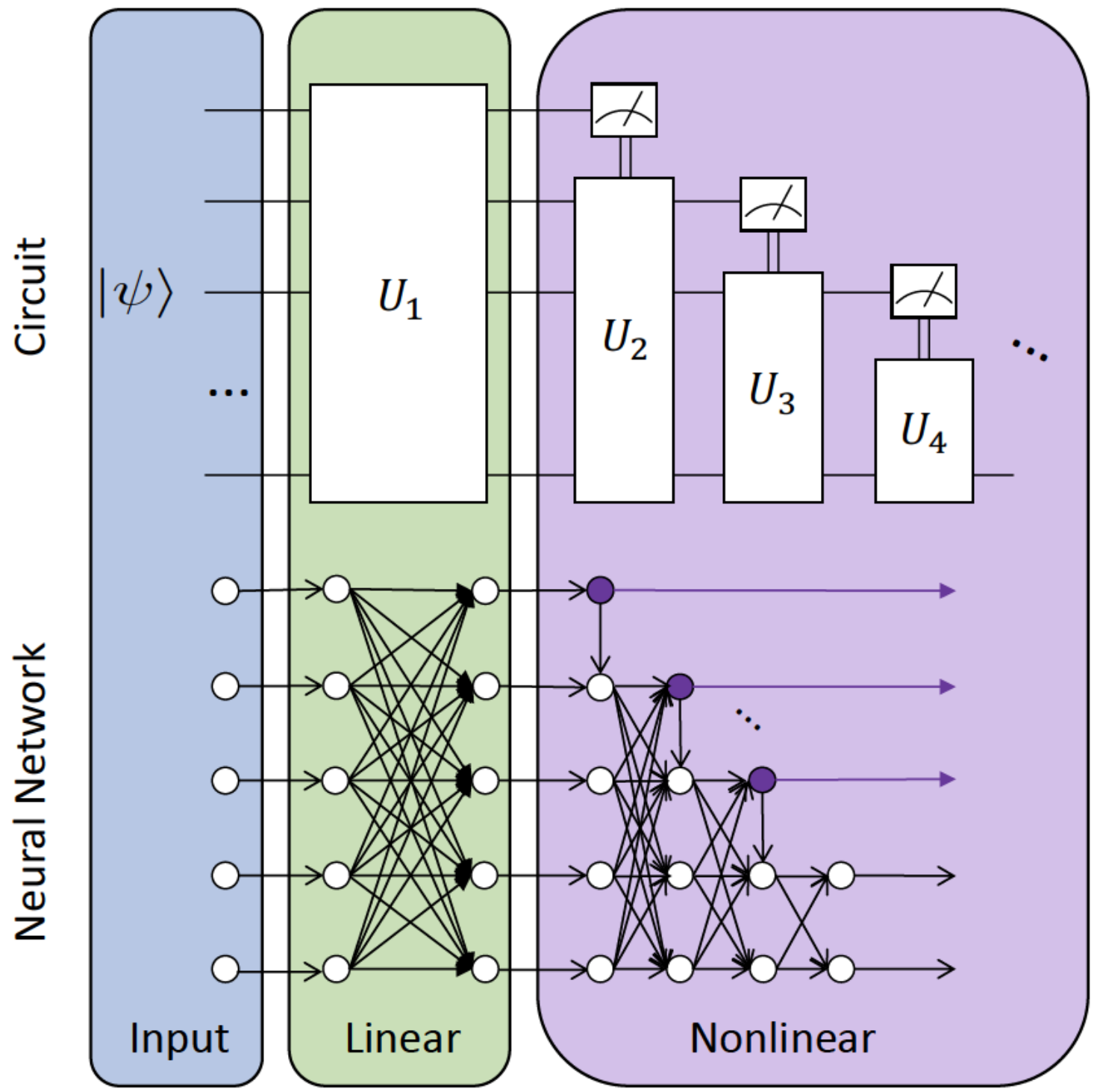}
\label{fig:qnn}
\caption{Universal quantum circuit learning construction can be viewed as quantum neural networks. The unitary evolution of quantum states could be considered as a layer of symmetric fully connected neurons. Multiple layers of generalized quantum measurements (POVMs) brings in sufficient non-linearities for performing classification or discrimination tasks.}
\end{figure}

In this work, we show that  universal shallow quantum circuits can be used as a universal discriminator for classification of quantum data from various different probability distributions approaching the optimal theoretical performance when such bounds are available. 
Our circuit topologies comprise gates from a universal gate set consisting of C-NOT and single-qubit gates, motivated by the fact that implementations of such are known for the currently most used experimental architectures. Furthermore our decompositions are nearly optimal in terms of the number of C-NOT gates which is an important feature for an implementation on near term devices. 
The quantum circuits we apply here can be viewed as a form of quantum neural networks with non-unitary layers, i.e., the generalized measurements, leading to sufficient non-linearities. A high-level description of a comparison of our quantum circuit learning and an analogy with a neural network structure is given in in Fig~\ref{fig:qnn}. Recently, unitary architectures have also been considered for the training of classical neural networks~\cite{arjovsky2016unitary,hyland2017learning}.

Specifically, we develop our quantum neutral networks based on optimal circuit decompositions for quantum channels, and directly adapt these channel decompositions to POVMs, conjecturing that our adapted decompositions for POVMs are also nearly optimal. We then train these circuits with a classical-quantum hybrid algorithm for the task of classifying different families of quantum data based on theoretical foundations of an important primitive known as \emph{quantum state discrimination}. Notably, unlike previous works, we focus on the generalisation ability of our circuit, i.e., we train the circuit on a specific range of the parameters with the goal of maximising its generalisation performance, hence considering a learning framework. This distinguishes our work from the pure optimisation problem for the state discrimination task, which is optimising the circuit to distinguish only a concrete set of states.

We show here with numerical simulations that the hybrid machine can learn a discrimination strategy either
in a minimal error setting, or in an unambiguous one, and observe a clear trade-off
between those two. One important feature of our discriminator quantum circuit model is that it uses a number of parameters that is polynomial in the size of the quantum states that are input into the circuit. In contrast, classical deep learning uses commonly a number of weights that is linear in the dimension of the input, however they will be exponential in the number of qubits for analysing quantum data.

The discriminative quantum neural networks introduced here could have broad range of applications. Quantum state discrimination by itself plays a key role in quantum information processing protocols and is used in quantum cryptography \cite{bennett1992quantum}, quantum cloning \cite{duan1998probabilistic}, quantum state separation and entanglement concentration \cite{chefles2000quantum,chefles2000quantum}. Our shallow quantum circuit learning could further be used to construct quantum repeaters and state purification units within quantum communication networks. It could also have wide range of applications, in quantum meteorology \cite{giovannetti2011advances} quantum sensing \cite{degen2017quantum}, quantum illumination \cite{lloyd2008enhanced}, and quantum imaging  \cite{chen2012optical} as a systematic way of engineering  structured receivers. In general, our discriminative networks could be used as a general quantum circuit verification units to examine the outputs of other shallow quantum circuits with possible applications to quantum versions of Generative Adversarial Networks (GANs) ~\cite{goodfellow2014generative}.

The paper is organized as follows. In
section~\ref{sec:QSD}, we first describe the family of quantum states that we aim to
discriminate, and introduce the optimization procedure that we use in the numerical
experiments. In section \ref{sec:train_with_exact}, we describe various
numerical experiments for training a quantum circuit where we use the exact
probabilities of the state measurements for our classical optimization of the parameters. We demonstrate that there exists a trade-off between the error probability $P_{err}$ and the probability of  inconclusive outcomes $P_{inc}$ when optimising our quantum circuits.
In section \ref{sec:use_measurement} we discuss the
number of measurement repetitions that we need in order to obtain a good estimate of the probabilities for the optimization task, i.e., optimal quantum state
discrimination. In
Appendix~\ref{sec:m_circ_parametrisation}, we describe the parametrization we
used for our quantum circuit, while Appendix~\ref{sec:adam_intro} provides a
brief introduction to the Adam stochastic optimization algorithm.

\section{Quantum state discrimination and classification \label{sec:QSD} }

Quantum state discrimination (QSD) is defined as follows: one is given a quantum device in an unknown state $\rho$ which is
believed to be from a family of non-orthogonal states $\{\rho_i\}$; the task is to design a
measurement to optimally identify $\rho$ \cite{Barnett2009,Bergou2007a}. Our fundamental inability to perfectly discriminate non-orthogonal quantum states is one of the key features of quantum information processing and quantum communication protocols ~\cite{Barnett2009}. For example, in a simple scenario in the context of quantum repeaters, when Alice wants to send a message to Bob, from previously agreed set, through a noisy quantum channel. Bob needs to discriminate the different messages within the background noise of the channel; that is he has to keep distilling the messages when they start overlapping with each other. Here we show that one could design a state discrimination measurement by training quantum circuits to accomplish this distillation. Ideally the training should be done in a secure and controlled environments to avoid tempering from malicious eavesdroppers.


Generally, there are two different strategies to cope with our inherent inability to perform  perfect QSD based on two different figures of merit. (a) \emph{Minimum-error state discrimination}: This is a deterministic strategy that we always make a judgment about the nature of the unknown state. The task is to minimize the inevitable errors in our classification. In this method the number of outcomes is equal to the number of possible states and we always can envision an optimal projective measurement according to Helstrom bound ~\cite{helstrom1969quantum}. However such optimal quantum measurements are very hard to find. (b) \emph{Unambiguous state discrimination}: This is a non-deterministic strategy. Here, we want to make sure with 100\% certainty that we have identified the correct state. Thus we need to introduce entropy sinks, here referred as "inconclusive outcomes", to attribute our lack of knowledge. Consequently, in the approach we need to rotate the states via unitary gates in a higher dimensional Hilbert space and then perform projective measurements. The number of outcomes will be always more than the number of possible input states. In other words, we always need to implement a POVM measurement. The maximum amount of information that can be extracted here is given by Holevo bound~\cite{holevo1973bounds}. As we will show here a pure unambiguous QSD is too costly with respect to inconclusive outcomes and one needs to allow for some small but non-zero errors to happen.  
We rely here on the formulation of the QSD problem as numerical optimization problem,
where one attempts to optimize one of the figures of merit~\cite{Eldar2003,Eldar2003a}.
However, rather than performing solely the optimisation, we use a machine learning approach to build a hybrid of the above two alternative strategies. We denote the probability of
making a successful judgment ($P_\text{suc}$), the probability of making a
wrong judgment ($P_\text{err}$), and the probability of making an inconclusive
judgment ($P_\text{inc}$), to evaluate a specific strategy. When the probably
of an inconclusive judgment is zero ($P_{inc}=0$) and the error probability
$P_{err}$ is minimized, the strategy becomes the standard minimal error
discrimination. When the error probability is zero, and the inconclusive rate
$P_{inc}$ is minimized, the strategy becomes the standard unambiguous
state discrimination.

In practice, we propose the usage of a classical computer to optimize
POVMs, implemented as parametrised quantum circuits, for state discrimination by changing the input quantum states
and parameters of a digital quantum processor. This quantum-classical hybrid method has the advantage of
delegating the expensive part of the optimization process, which is the
evolution of quantum states, to a quantum processor and can be considered as quantum neural networks. The reason is the close mathematical relation between the unitary circuit structures  and classical unitary-weight neural network structures~\cite{schuld2018circuit,arjovsky2016unitary,hyland2017learning}. It has been shown that the unitarity of the layers indeed is optimal since it can avoid the exploding or vanishing gradient problem if the activation function is chosen adequately, e.g., a ReLu~\cite{arjovsky2016unitary}. Since quantum circuits do not include any activation functions between the layers, unless a projective measurement is performed, this advantage is immediately transferable to the case of quantum circuit training. As pointed out already in~\cite{schuld2018circuit}, it has further been shown that unitary weight matrices further allow for gradient descent methods to converge independently of the circuit depth~\cite{saxe2013exact}, which is important for the training of circuits with larger circuit depth.

Here we treat the general problem of state discrimination of an the ensemble of families of pure states. In this setting, each state in the ensemble is be drawn from a family of states, each parametrised by a specific distribution. Concretely the input is then given by
\begin{equation}
\rho = \sum_{i=1}^m \lambda_i \ket{\psi_i( \alpha_{i})}\bra{\psi_i(\alpha_{i})}, \quad \sum_{i=1}^m \lambda_i =1,
\end{equation}
where any $\alpha_k$ can be a distinct discrete classical probability distribution, and we use the short notation $$\ket{\psi_i(\alpha_i)} := \ket{\psi_i(a_i)}_{a_i \sim \alpha_i} = \sum_j \beta_j(a_i) \ket{j},$$ where $a \sim \alpha_i$ means that $a$ is drawn from the distribution $\alpha_i$, and $\beta(a)$ indicates some distribution for the amplitude parametrised by $a$.
In particular this means that we perform state discrimination on the ensemble of parametrised pure states $\{ \ket{\psi_i(a_i)} , \alpha_i \}_{i=1}^m$ where the parameters $a_i$ that characterise the amplitudes of $\ket{\psi_i(a_i)}$ are distributed according to $p(a_i)\sim \alpha_i$. We hence draw the $a_i$ from the according probability distribution $\alpha_i$ and then use $a_i$ to specify the amplitude distribution, i.e.,  $\ket{\psi_i} = \sum_{j=1}^N \beta_i (a) \ket{i}$, where the family of states $\ket{\psi_i(\alpha_i)}$ is then used as input into the circuit with some probability $\lambda_i$. 


We now address a canonical example of
discriminating among two families of non-orthogonal quantum states over the Hilbert
space of a two-qubit system. One of our inputs belong to a family of pure states $\psi_1(a)$ parametrised by a real number $a\in[0,1]$:

\begin{equation}
  \label{eq:psi_1}
  \psi_1(a) = \left( \sqrt{1-a^2}, 0, a, 0 \right)
\end{equation}

The second input class consists of a family of mixed states represented by two, up to a sign, equal pure states $\psi_{2/3}(b)$ which are parametrised by a real number $b\in[0,1]$:

\begin{align}
  \label{eq:psi_23_b}
  \psi_{2/3} &= \left( 0, \pm \sqrt{1-b^2}, b, 0 \right)
\end{align}

For simplicity, we first focus on the task of supervised learning of states that are parametrised by $a\in[0,1]$, i.e., the first family of quantum states and using a fixed parameter $b= \frac{1}{\sqrt{2}}$ for the second set. This allows a direct comparison to theoretical and prior optical realization of quantum state discrimination and the related task of quantum state filtering~\cite{Mohseni2004}. We therefore will work here with the states of the form:

\begin{align}
  \label{eq:psi_23}
  \psi_1(a) &= \left(\sqrt{1-a^2}, 0, a, 0 \right),\quad \text{and}\nonumber \\
  \psi_{2/3} &= \left(0, \pm \frac{1}{\sqrt{2}}, \frac{1}{\sqrt{2}}, 0 \right).
\end{align}

For comparison of~\cite{Mohseni2004} we set $\lambda_1=1/3$ and $\lambda_2=2/3$. The overlap (fidelity) between $\psi_1(a)$ and $\psi_{2/3}$ is given by $a/\sqrt{2}$. This means roughly, that the task of discriminating $\psi_1(a)$ and $\psi_{2/3}$ becomes increasingly harder with increasing $a$. The unambiguous discrimination of these two families has in particular been demonstrated experimentally for values $a=0.25$ and $a=0.5$ by Ref. ~\cite{Mohseni2004}, which we will use to benchmark our own results.

In this paper, we use a quantum-classical hybrid scheme, in which the
(simulated) quantum computer is used as a subroutine for the optimisation task which is called from a classical device. The classical machine thereby controls the input states to the
quantum circuit, assessed the output and finally optimizes parameters of the quantum circuits depending on the results of this subroutine.
Our quantum circuit is implemented as a general POVM with $4$
possible measurement outcomes (Eq.~\ref{eq:general-measurement-circuit}).
The parametrisation of the circuit is described in detail in Appendix~\ref{sec:m_circ_parametrisation}.
We hence use $2$ (ancilla) qubits for the measurements, and denote
these measurement outcomes as $m_{i_2i_1}$, where
$i_1, i_2 \in \{0,1\}$ are the measurement outcomes of first and second qubit respectively. Since we have four possible outcomes but only three different input
signals, i.e., $\psi_1(a)$, $\psi_{2/3}$, and the inconclusiveness result, we arbitrarily choose a
correspondence between outcomes and signals. Here in particular we assign $m_{00}$ or $m_{10}$ to the input $\psi_1(a)$, while a measurement outcome of $m_{01}$ is assigned to the input $\psi_{2/3}$, and $m_{11}$ to the inconclusive measurement.
With these assumptions, the various probabilities
mentioned before ($P_{suc}$, $P_{err}$, and $P_{inc}$) can be defined in a
natural way. A simplified version of the circuit we use for the discrimination task is given below.

\begin{equation}
\label{eq:general-measurement-circuit}
   \Qcircuit @C=1em @R=.7em {
     \lstick{\ket{0}}    & {/^2}\qw & \multigate{1}{V} & \meter & \rstick{M} \\
     \lstick{\ket{\psi}} & {/^2}\qw & \ghost{V}        & \qw
}
\end{equation}
\smallskip

In construction of shallow circuits, since C-NOT gates are usually much more prone to errors than single-qubit gates, an ideal decomposition minimizes the number of C-NOTs required for a particular task. The problem of finding near optimal (in terms of the required number of C-NOT gates) circuit topologies for quantum channels
from $m$ to $n$ qubits was considered in~\cite{Iten2015,Iten2016}.
We adapt here the construction given therein to find circuit topologies for POVMs. The procedure is as follows.
Since all possible POVMs on $m$ input qubits can be represented by a quantum channel
from $m$ to $n$ qubits, we can find the low-cost circuit topology for POVMs, which is universal, by applying a QR decomposition and a Cosine-Sine decomposition of the quantum channel, similarly to~\cite{Iten2015,Iten2016} (Appendix \ref{sec:m_circ_parametrisation}).
The high level idea for proving the near optimality of the channel decomposition is then
based on a parameter counting argument. One inspects the manifold of the quantum channel and determines its dimensionality, which gives
a lower bound for the number of parameters and hence for the number of single-qubit gates which need to be introduced in the circuit topology. 
If this number of parameters is not introduced, one can only construct a set of measure zero in the manifold of all channels with the given circuit topology.

Next, we define our theoretically motived cost function that is used for our optimisation task throughout the paper. Denote $S_i$ the quantum states for different families of distribution acquired to train the circuit, which could be viewed as a specific sampling of the distributions $\alpha_i$. The cost function is then defined as:

\begin{align}
	\label{eq:J_1}
	J_1 =&
	\sum_i \frac{1}{|S_i|} \sum_{a_i\in S_i} \left( 1-P_\text{suc}(\psi_i(a_i)) \right)
	\nonumber \\
	& + \alpha_\text{err} \sum_i \frac{1}{|S_i|} \sum_{a_i\in S_i}
	    P_\text{err}(\psi_i(a_i))  \nonumber \\
	& + \alpha_\text{inc} \sum_i \frac{1}{|S_i|} \sum_{a_i\in S_i}
	    P_\text{inc}(\psi_i(a_i)),
\end{align}
where $|S_i|$ is the number of samples in the training sets $S_i$, $\alpha_\text{err}$ is the penalty for making
errors, and $\alpha_\text{inc}$ is the penalty for giving inconclusive outcomes.
$P_\text{suc}(\psi)/P_\text{err}(\psi)/P_\text{inc}(\psi)$ are the probabilities
of giving a correct/erroneous/inconclusive measurement outcome for a input wave function
$\psi$. This cost function reflects three things: the award given to a
successful discrimination of the states, the penalty (controlled by $\alpha_{err}$) for a wrong discrimination, and the penalty (controlled by $\alpha_{inc}$) of an
inconclusive result. The probabilities ${P_{suc/err/inc}(\psi)}$ are directly accessible
in the numerical simulations, and can in practical experiments be inferred through repeated measurements on $\psi_1$, up to some precision. Eq.~\ref{eq:J_1}
 hence measures the performance of the quantum circuit~\ref{eq:general-measurement-circuit}, which we use in the following for the state discrimination task.

Since the families of quantum states we aim to discriminate consist of a set of infinitely many wave functions ($a$ is a real number),
we adopt a strategy which is common in machine learning.
We sample a set of values $a$ from the range $[0,1]$ and divide the resulting set into a training and a test set.
The training set (denoted $S_\text{train}$) is then used to
calculate the cost function and to optimise the quantum circuit during the training
phase, while the test set (denoted $S_\text{test}$) is used
for the verification of the performance of the trained model, i.e., in our case the quantum discrimination circuit.
The concrete optimisation of the circuit with the classical machine was done by minimising our cost function for the training set using Adam, where the gradients are calculated using forward differences.


%% file: sections/results.tex
\section{Training on simulated quantum computers}
\label{sec:train_with_exact}

In this section, we describe several numerical experiments performed on
classical hardware using the exact probabilities as defined
above. The extension of this approach is then
treated in Section~\ref{sec:use_measurement}. In all the numerical experiments, we used the three values $P_\text{suc}$, $P_\text{err}$
and $P_\text{inc}$ as defined in equation \ref{eq:p_suc_p_err_p_inc} to summarise and compare the training results. We found empirically that saddle points often exists in training and hence stochastic gradient descent does not perform well. This motivates the use of Adam over stochastic gradient descent throughout the experiments. We will later compare the performances of various different stochastic optimization techniques. 

\begin{subequations}
\label{eq:p_suc_p_err_p_inc}
\begin{align}
  P_\text{suc} &= \sum_i \lambda_i \frac{1}{|S_i|} \sum_{a_i\in S_i} P_\text{suc}\left(\psi_i(a_i)\right),  \label{eq:p_suc}
   \\
  P_\text{err} &= \sum_i \lambda_i \frac{1}{|S_i|} \sum_{a_i\in S_i}
  P_\text{err}\left(\psi_i(a_i)\right), \label{eq:p_err}
   \\
  P_\text{inc} &= \sum_i \lambda_i \frac{1}{|S_i|} \sum_{a_i\in S_i}
  P_\text{inc}\left(\psi_i(a_i)\right), \label{eq:p_inc}
\end{align}
\end{subequations}
where $ S_i$ here is the training or test set.

\begin{figure*}
  \centering
  \subfloat[$a_0=0.25$, $P_\text{suc}=0.84$]{
    \includegraphics[width=0.45\textwidth]{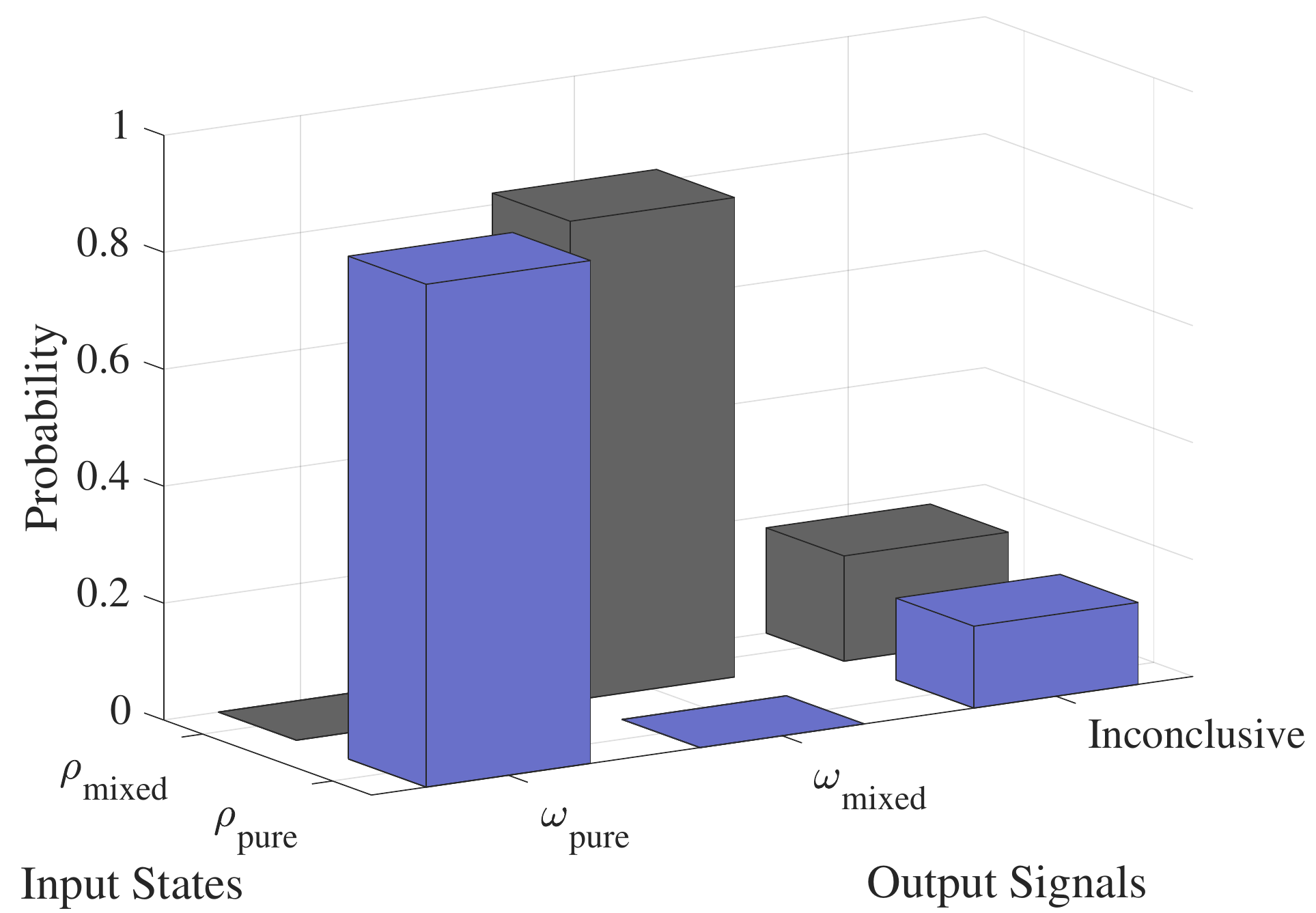}
  }
  \subfloat[$a_0=0.50$, $P_\text{suc}=0.59$]{
    \includegraphics[width=0.45\textwidth]{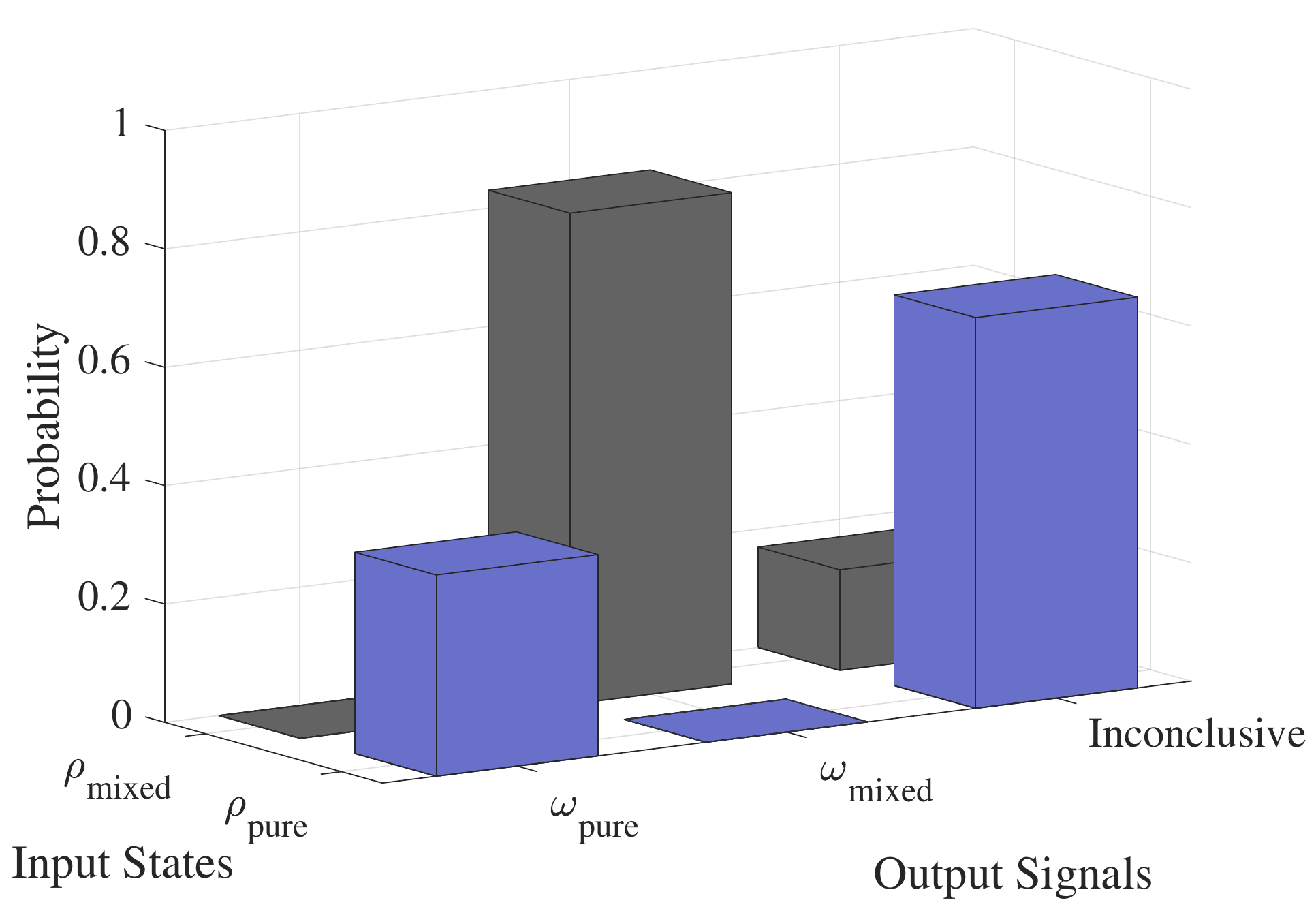}
  }
  \caption{Performance for unambiguous discrimination with training data centred at
  $a_0$ with standard deviation $0.01$. The
    penalty parameters are set to be $(\alpha_{err},\alpha_{inc})=(25,2)$.
    The trained circuit unambiguously discriminates the pure states
    $\rho_\text{pure}=\ketbra{\psi_1(a_0)}$ from the mixed states
    $\rho_\text{mixed} = 1/2(\ketbra{\psi_2}+\ketbra{\psi_3})$.
    The values
    shown are the exact probabilities for the
    given specific measurement outcome, and are averaged over $50$ runs.
    For reference, we provide the theoretically optimal success rate of $0.833$ for
    $a_0=0.25$ and $0.666$ for $a_0=0.50$ from Mohseni, et. al., \cite{Mohseni2004}.}
  \label{fig:acen_perf}
\end{figure*}

\begin{figure*}
	\centering
	\includegraphics[width=0.7\textwidth]{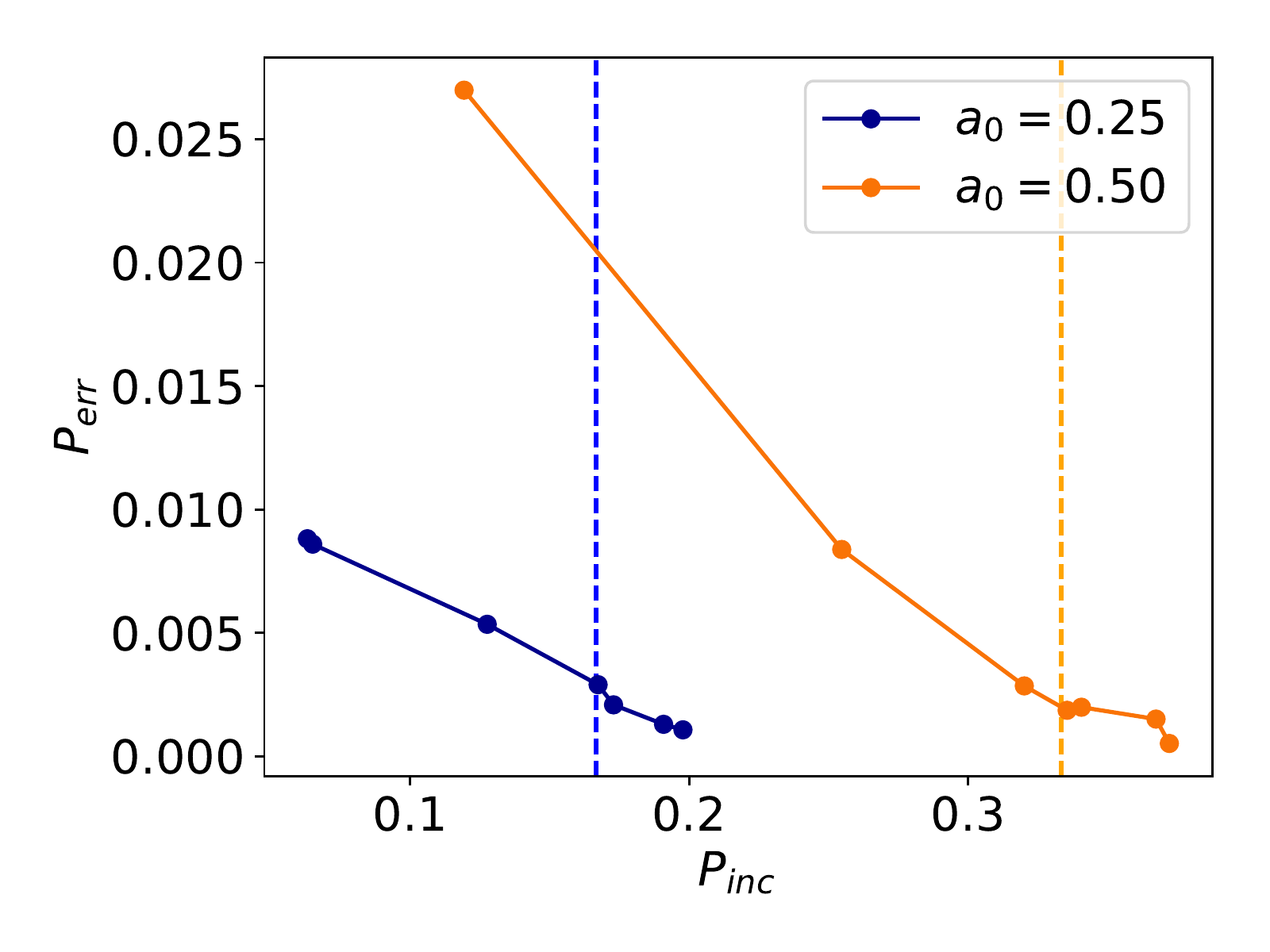}
  \caption{Trade-off between the error probability and the inconclusiveness for
    training data centred at $a_0$ with standard deviation $0.01$. The
    lines represent averaged quantities of $P_{err}$ and $P_{suc}$ which we obtain by increasing
    $\alpha_{err}$ from $10,15,20,\cdots$ to $40$, with $\alpha_{inc}$ fixed at
    $2$. Bars represent the standard deviation. The dashed vertical line represents
    the theoretical minimal inconclusive rate for unambiguous state discrimination~\cite{Mohseni2004}.
    Although the training does not reach the minimal inconclusiveness rate
    when we require $P_\text{err} \approx 0$, we are able to obtain much smaller
    $P_\text{inc}$ by accepting small non-zero $P_\text{err}$. This trade-off between
    $P_\text{err}$ and $P_\text{inc}$ can be useful in realistic applications.
    The standard deviation of $P_\text{err}$ are estimated to be 0.001 to 0.01
    and 0.04 to 0.12 for $P_\text{inc}$.
  }
	\label{fig:perrinctradea025050}
\end{figure*}

\paragraph{Discriminate an input from a normal distribution centred at $a_0\in[0,1]$.---}
In this experiment, we optimise the cost function $J_1$ with $a \in
S_\text{train}$ centred around $a_0$\footnote{The centred data was chosen from
  100 points that were i.i.d., randomly sampled from a normal distribution with
  mean $\mu=a_0$ and variance $\sigma=0.01$. The normal distribution was
  truncated so that $0\leq a \leq 1$.}
, which allows for a direct comparison of our trained circuit with the results
for unambiguous state  discrimination of \cite{Mohseni2004}\footnote{Note that
this problem is commonly denoted as state filtering (SF).}.
We train our model using Adam
\footnote{The Adam used a stochastic gradient taking $50$ sub-samples of $a$ in
  the training set. Parameters for Adam are
  $\beta_1=0.9,\beta_2=0.999,\hat{\varepsilon}=10^{-8}$ and learning rate
  $0.001$, as in Appendix~\ref{sec:adam_intro}.  The gradient was calculated
  using the forward differences formula with step size $10^{-6}$.
} and test it on the test set $S_\text{test}=\{a_0\}$.
In repeated experiments, we did not reach on average the theoretical minimum inconclusiveness $P_\text{inc}$ when we use a high penalty to force the error probability $P_\text{err}$ to be close to zero. But since the training took data with $a$ centred at, but not fixed at the point $a_0$, we should not expect the trained circuit to obtain exactly the same minimal value.
Also, we are able to obtain a $P_\text{inc}$ which is close the the theoretical minimum,
if we allow a small, but non-zero error probability. This demonstrates that a less strict constraint on
$P_\text{err}$ allows for a much lower inconclusiveness rate, i.e. there is a
trade-off between the error rate ($P_{err}$) and the inconclusive rate ($P_{inc}$) for
this problem during the training process (c.f.~Fig.~\ref{fig:perrinctradea025050}).
To allow a visual comparison with Fig.~2 in \cite{Mohseni2004}, we first show
the averaged performance for our circuit for $(\alpha_{err},
\alpha_{inc})=(25,2)$ in Fig.~\ref{fig:acen_perf} and then the resulting trade-off in~Fig.~\ref{fig:perrinctradea025050}.

\begin{figure*}
	\centering
	\subfloat[{Unambiguous discrimination.\newline
		(Tested on $a\in [0,1]$.)}]{
		\includegraphics[width=0.45\linewidth]{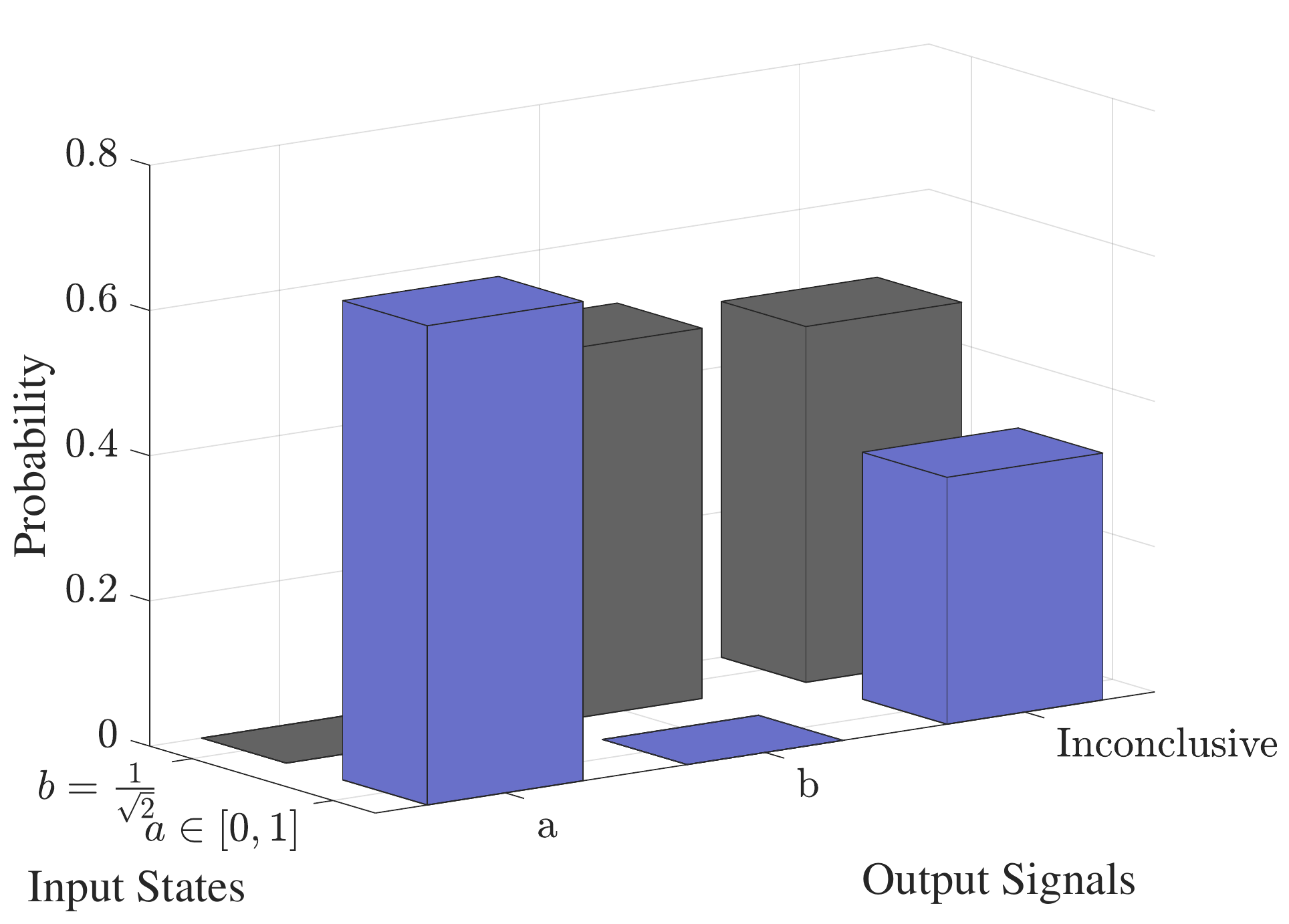}
	}
	\subfloat[{Unambiguous discrimination.
		\newline (Tested on $a\in[0.9,1]$.)}]{
		\includegraphics[width=0.45\linewidth]{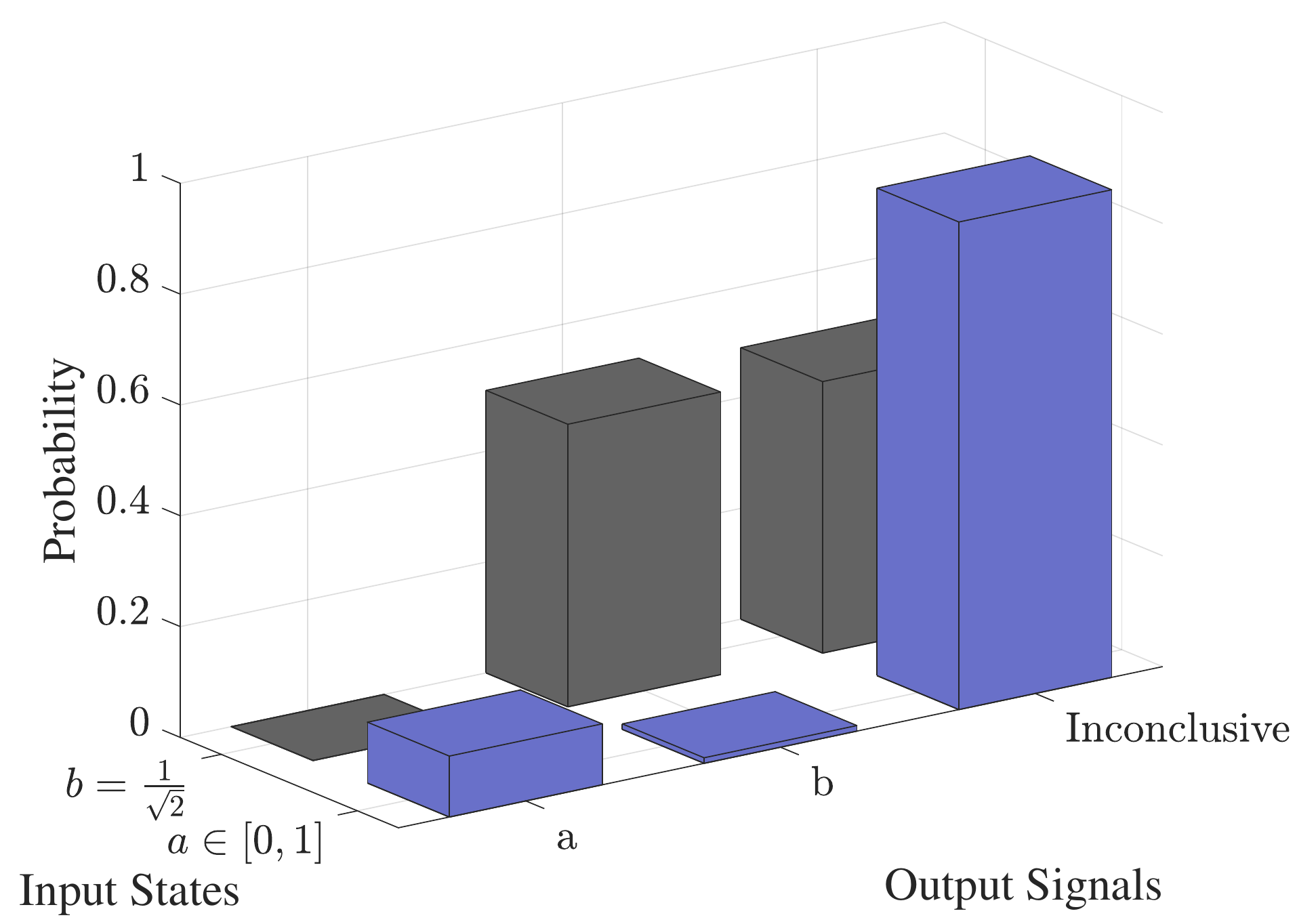}
	}
	\\
	\subfloat[{Minimal Error Discrimination.
		\newline (Tested on $a\in[0,1]$.)}]{
		\includegraphics[width=0.45\linewidth]{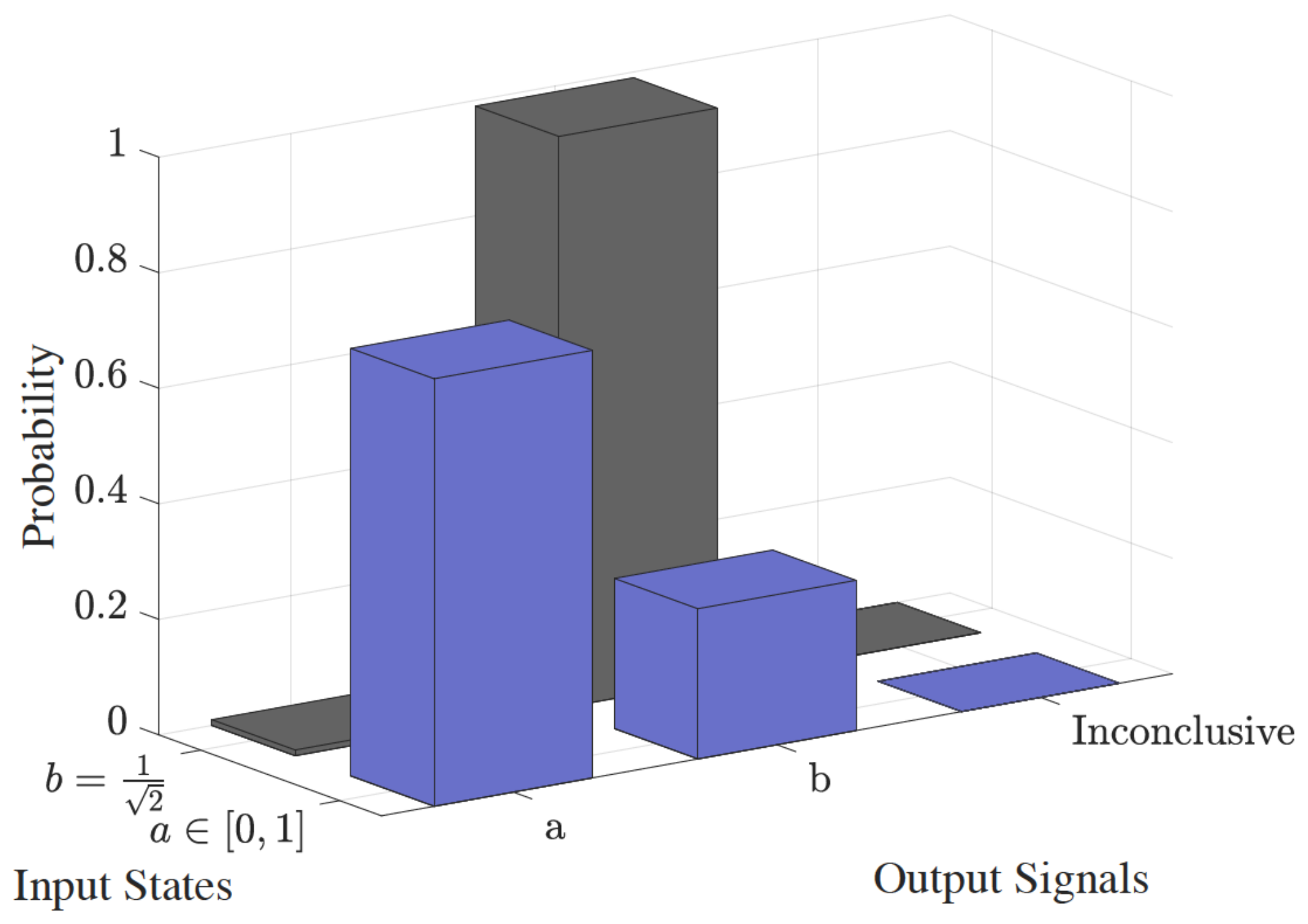}
	}
	\subfloat[{Minimal Error Discrimination.
		\newline (Tested on $a\in[0.9,1]$.)}]{
		\includegraphics[width=0.45\linewidth]{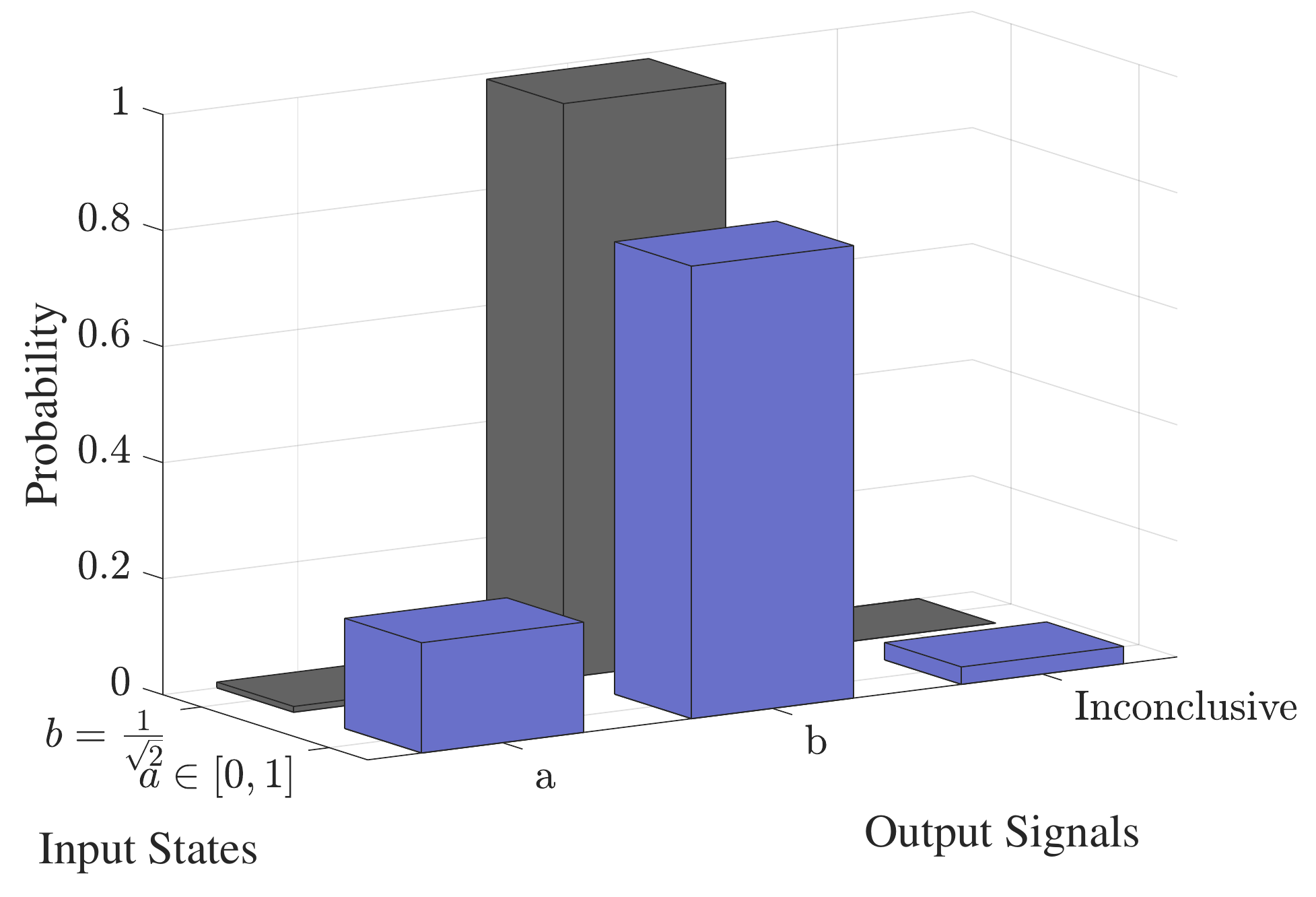}
	}

	\caption{Performance of learned quantum circuits: The values shown
    above are the exact probabilities for a given specific measurement outcome
		averaged over $50$ repeated runs. The probability for $\psi_1(a)$ is
		averaged over all $a\in S_\text{test}$. Here we find two types of
		discrimination strategy. When $\alpha_{err}>\alpha_{inc}$, (Fig.~(a) and Fig.~(b), where $(\alpha_\text{err},\alpha_\text{inc})=(20,2)$), we
		obtain unambiguous discrimination with zero error rate. When
		$\alpha_{inc}>\alpha_{err}$ (Fig.~(c) and Fig.~(d), where $(\alpha_\text{err},\alpha_\text{inc})=(5,20)$), we obtain a minimal
		error discrimination with zero inconclusiveness rate. By comparing Fig.~(a) with Fig.~(b)
		(and (c) with (d)), one notices the general feature that the degree of non-orthogonality of the wave function determines the hardness.
	}
	\label{fig:ud_me_example}
\end{figure*}

\begin{figure*}
  \centering
  \subfloat[$P_\text{err}$]{
  	\includegraphics[width=0.45\textwidth]{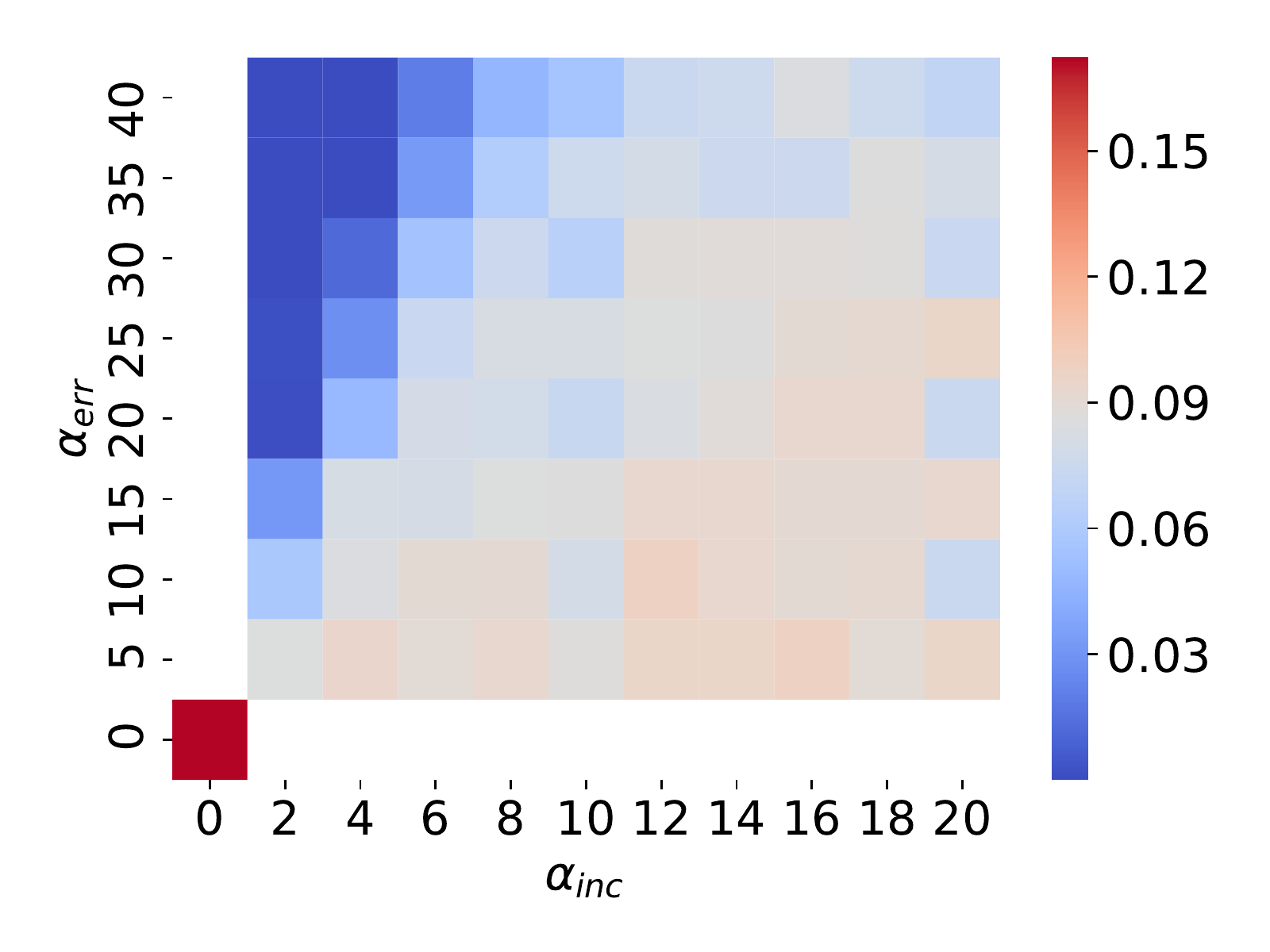}
  }
  \subfloat[$P_\text{inc}$]{
    \includegraphics[width=0.45\textwidth]{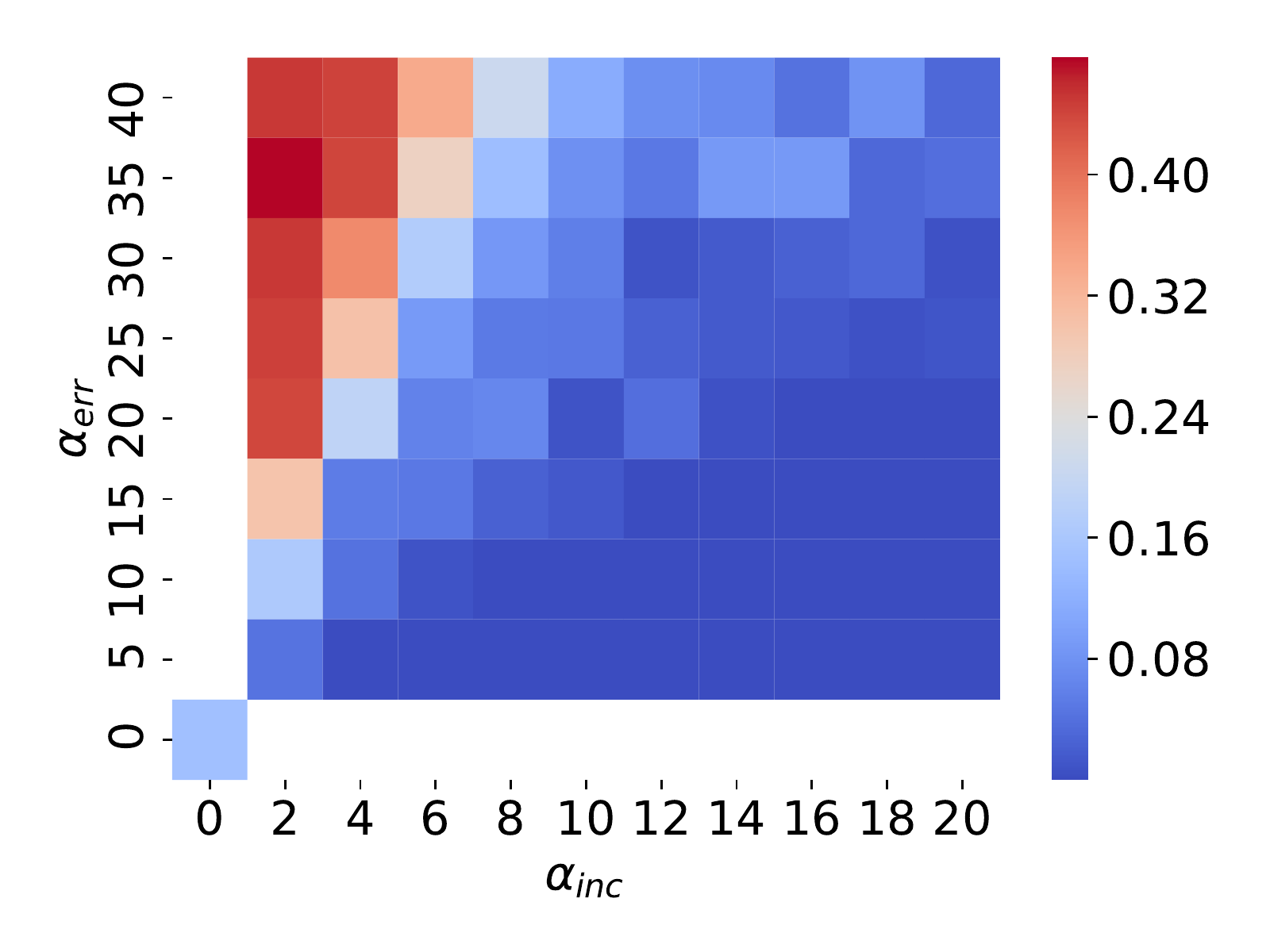}
  }
  \\
  \subfloat[$P_\text{suc}$]{
    \includegraphics[width=0.45\textwidth]{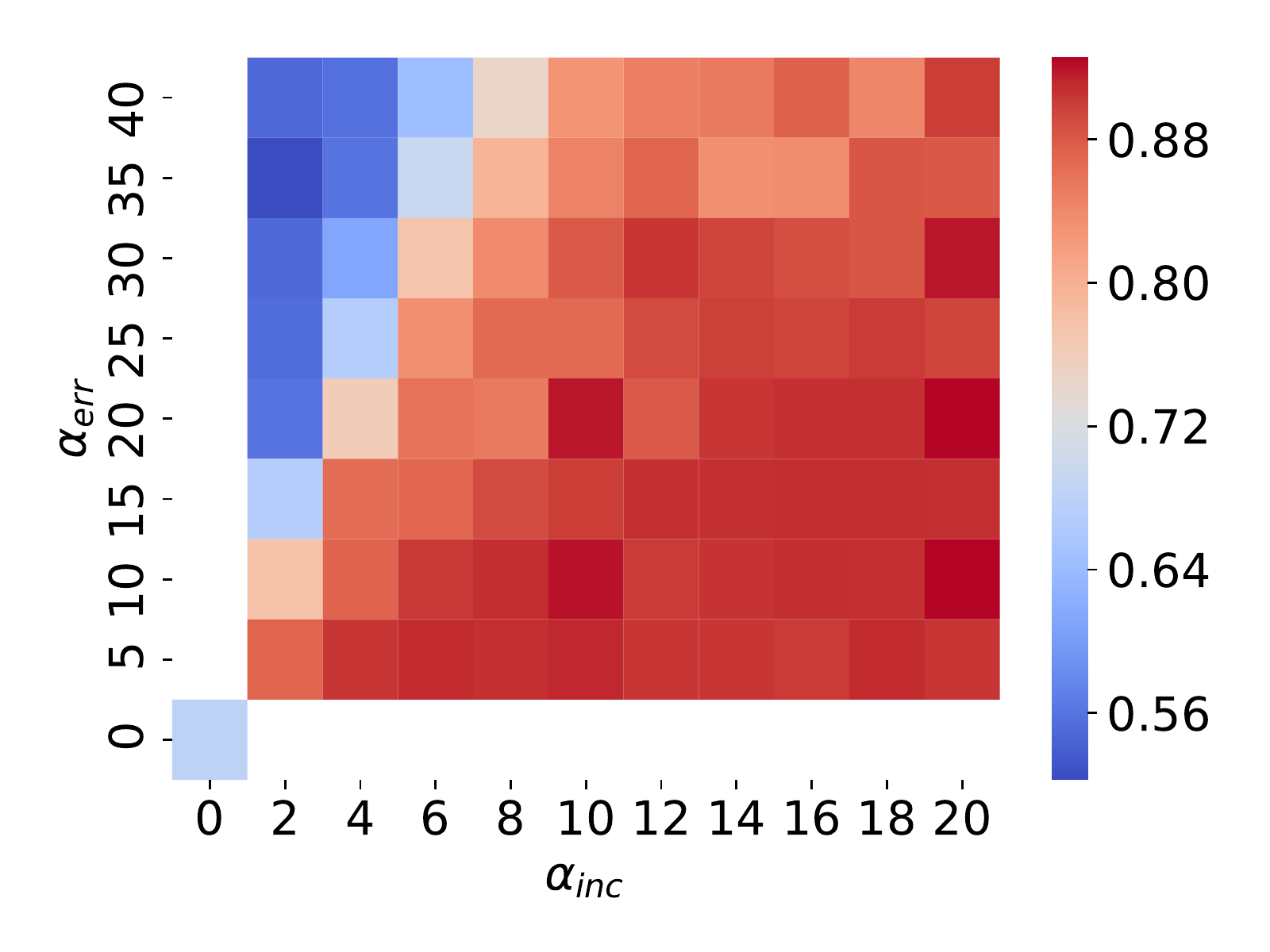}
  }
  \subfloat[$P_\text{suc}$ SD]{
  	\includegraphics[width=0.45\textwidth]{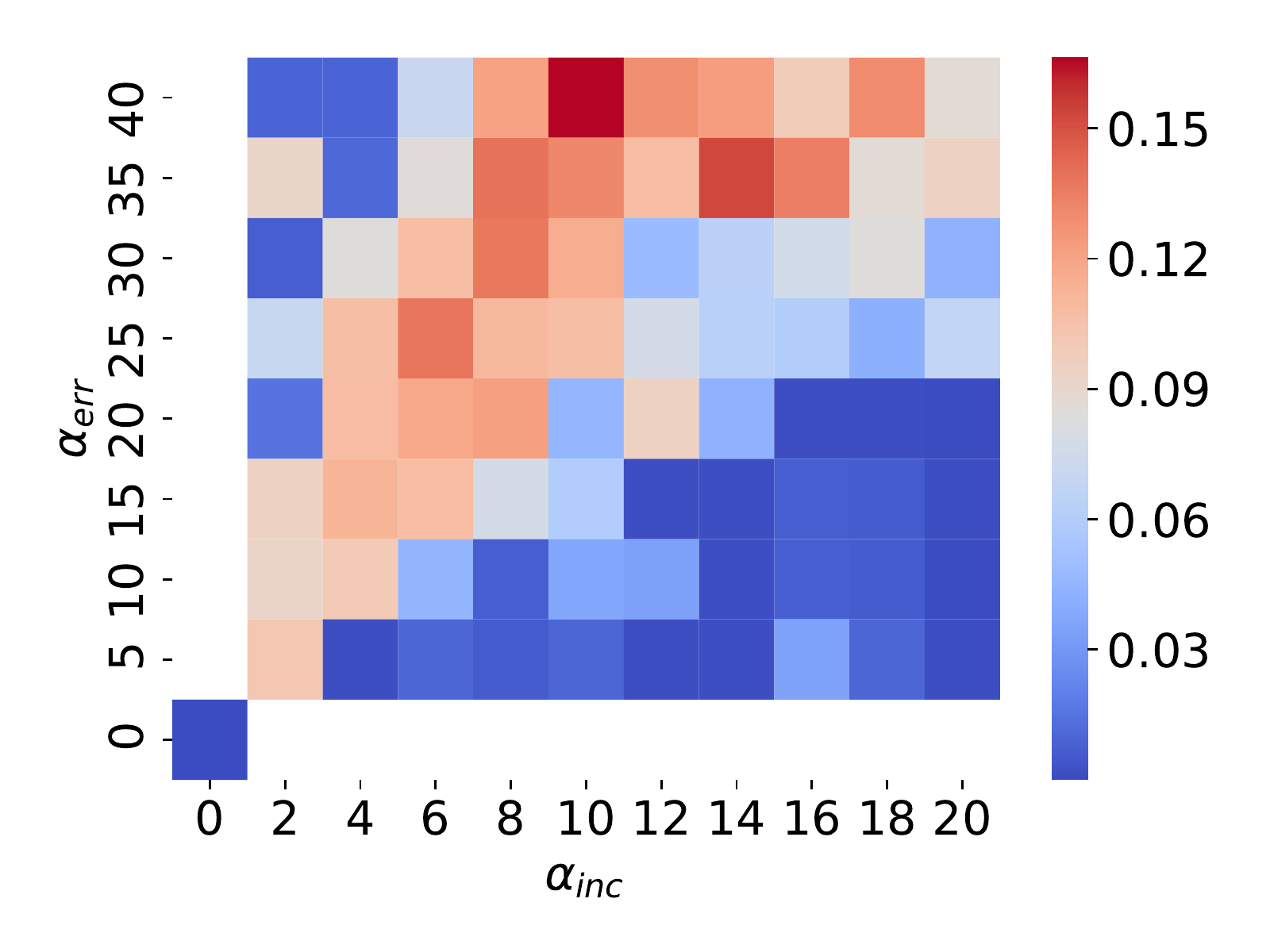}
  }
  \caption{
    Here, Fig.~(a)-(c) show the gradual transition from the unambiguous
    discrimination (near-zero error rate) to a minimal error discrimination
    (near-zero inconclusiveness) with different parameters $\alpha_\text{err}$
    and $\alpha_\text{inc}$. The bottom right regions in all plots are mostly
    uniform, which shows that it is much harder to obtain a small inconclusiveness
    rate than a small error rate. Compared with the point
    $\alpha_\text{err}=\alpha_\text{inc}=0$, the added penalties improve
    the success probability or the inconclusiveness respectively. The data was tested on
    $a\in[0,1]$, and averaged over $50$ runs. The standard deviation
    for $P_\text{suc}$ is shown in Fig.~(d). With an increasing standard
    deviation (closer to the diagonal line), the result becomes increasingly
    unstable when the two penalties ($\alpha_\text{err}$ and
    $\alpha_\text{inc}$) are comparable in magnitude. The standard deviation
  for other values shows the same pattern as for $P_\text{suc}$.}
  \label{fig:perr_inc_diff_penalty}
\end{figure*}

\paragraph{Discriminate among all data $a\in [0,1]$.---}
Here we aim to discriminate the inputs from two classes of states where
the second class are wave functions coming from the family $\psi_1(a)$ with in $a\in[0,1]$.
We train and test the circuits with $S_\text{train}$ and $S_\text{test}$
which are two random subsets sampled from the range $[0,1]$\footnote{
  The $J_1$ was optimised with $S_\text{train}$ chosen to be $100$
  evenly spaced points from the range $[0, 1]$. The test set $S_\text{test}$ was
  chosen to be $150$ i.i.d., randomly chosen points in $[0,1]$
}. Inspired by the trade-off found in previous results for discriminating a
single data point, we train the cost function $J_1$ for different choices of the parameters
$(\alpha_{err}, \alpha_{inc})$ using Adam\footnote{We use as input parameter
  $\beta_1=0.9,\beta_2=0.999,\hat{\varepsilon}=10^{-8}$ and a learning rate of
  $0.001$ for Adam (c.f.~Appendix~\ref{sec:adam_intro}). The gradient is
  calculated stochastically using  $50$ sub-samples chosen i.i.d.\ at random from
  $S_\text{train}$. The step size for the numerical calculation of the gradient is
  chosen to be $10^{-3}$ for the forward differences formula. All data is
  obtained after $5000$ iterations of Adam.
}. Our results show that in comparison with the case of zero penalties
($\alpha_\text{err}~=~\alpha_\text{inc}~=~0$), the penalties act as a form
of regularisation and can be adjusted to give a higher
success probability or a lower inconclusiveness rate for the final model. Furthermore, we
observe a gradual transition from unambiguous discrimination
(characterised by near-zero error probability) to minimal error discrimination
(characterised by near-zero inconclusiveness) with varying penalties
(c.f.~Fig.~\ref{fig:perr_inc_diff_penalty}). The results for the trained circuit
for specific pairs of $\alpha_\text{err}$ and $\alpha_\text{inc}$, after being
fine-tuned to closely resemble both the unambiguous discrimination strategy and
the minimal error discrimination strategy, are shown in
Fig.~\ref{fig:ud_me_example}.


\begin{figure*}
    \centering
      \includegraphics[width=0.6\linewidth]{%
      	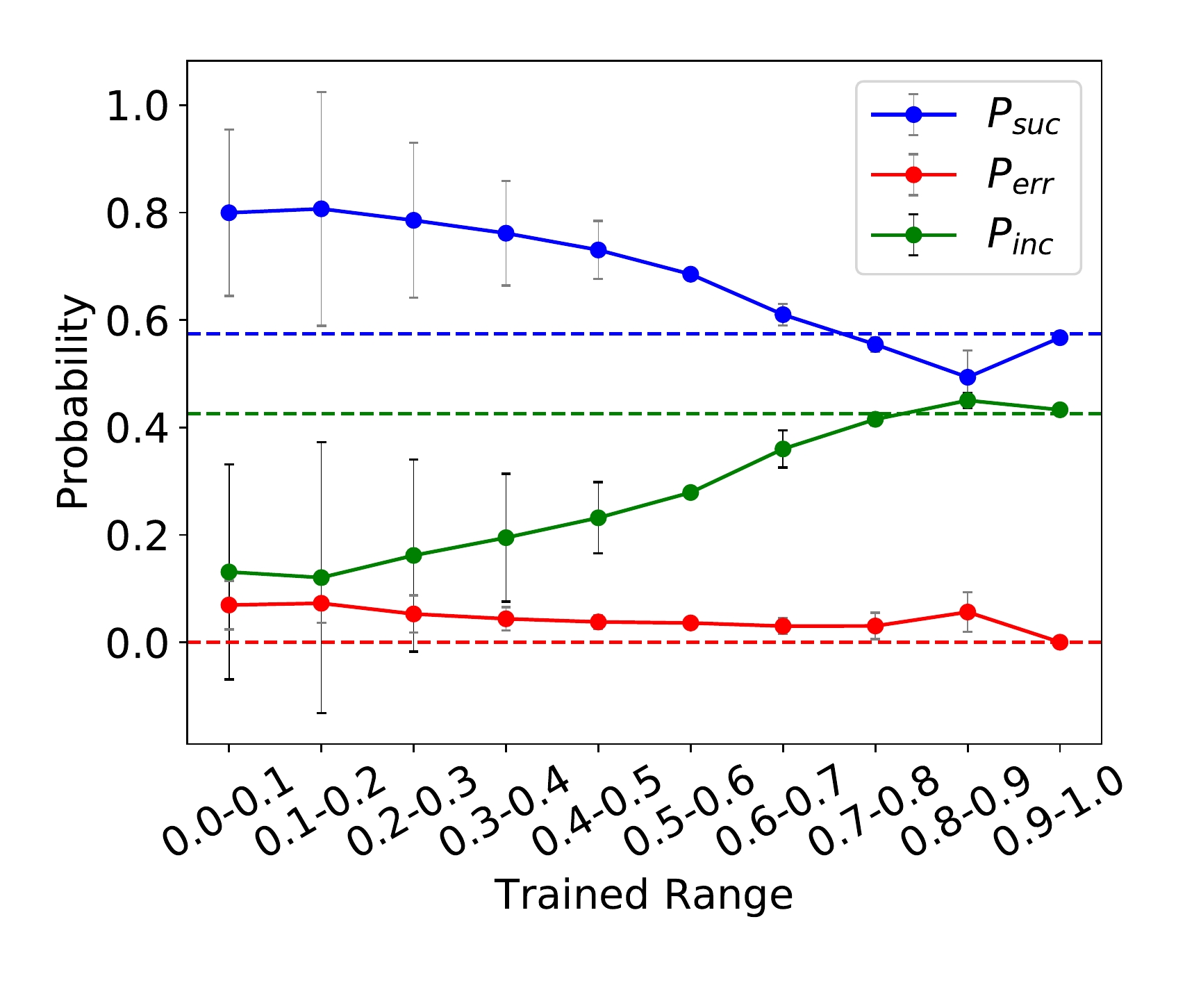}
    \caption{Performance of a circuit trained on a small data range, while
      tested on the full range of $a\in[0,1]$. The dashed horizontal line gives as reference
      the results for a circuit which is trained on the complete range $a\in[0,1]$.
      The circuits trained on a small range $a$ close to $1$, shows a performance that is similar to
      a circuits trained on the whole range of $a\in[0,1]$. We therefore conclude, that the training is
      highly dominated by the wave functions $\psi_1(a)$ with $a$ close to $1$.
      This reflects the fact that with increasing fidelity
      ($F(\psi_1(a),\psi_{2/3})\propto a$), two quantum states are generally
      harder to discriminate. The test data was averaged over $10$ repeated
      runs with parameters $\alpha_\text{err}=40$ and $\alpha_\text{inc}=4$. The
      bars indicates the standard deviations.
    }
  \label{fig:generalisation}
\end{figure*}

\paragraph{Generalisability of training data.---}
On the other hand, we hypothesize that the fidelity is a good indicator of the
generalisation ability of the trained model.
To test this hypothesis, we train the
circuit with $a$ restricted to a subsets of $[0,1]$, where $a$ is chosen close to $1$ (recalling that the
fidelity between $\psi_1(a)$ and $\psi_{2/3}$ is proportional to $a$), and then
test the performance of the trained circuit on the full range of $a\in[0,1]$.
The parameters for the training are the same as in the previous experiment with
centred data. We find for the unambiguous discrimination case (i.e., large penalty
for errors), that the training result is dominated by wave functions
$\psi_1(a)$ with $a$ close to $1$(c.f.~Fig.~\ref{fig:generalisation}). This
reflects the difficulty of distinguishing $\psi_1(a)$ from $\psi_{2/3}$ if $a$ is close to $1$.

\begin{figure*}
 \centering
 \subfloat[{Averaged fidelity between wavefunctions, with axis labelling the source of data}]{
   \includegraphics[width=0.45\linewidth]{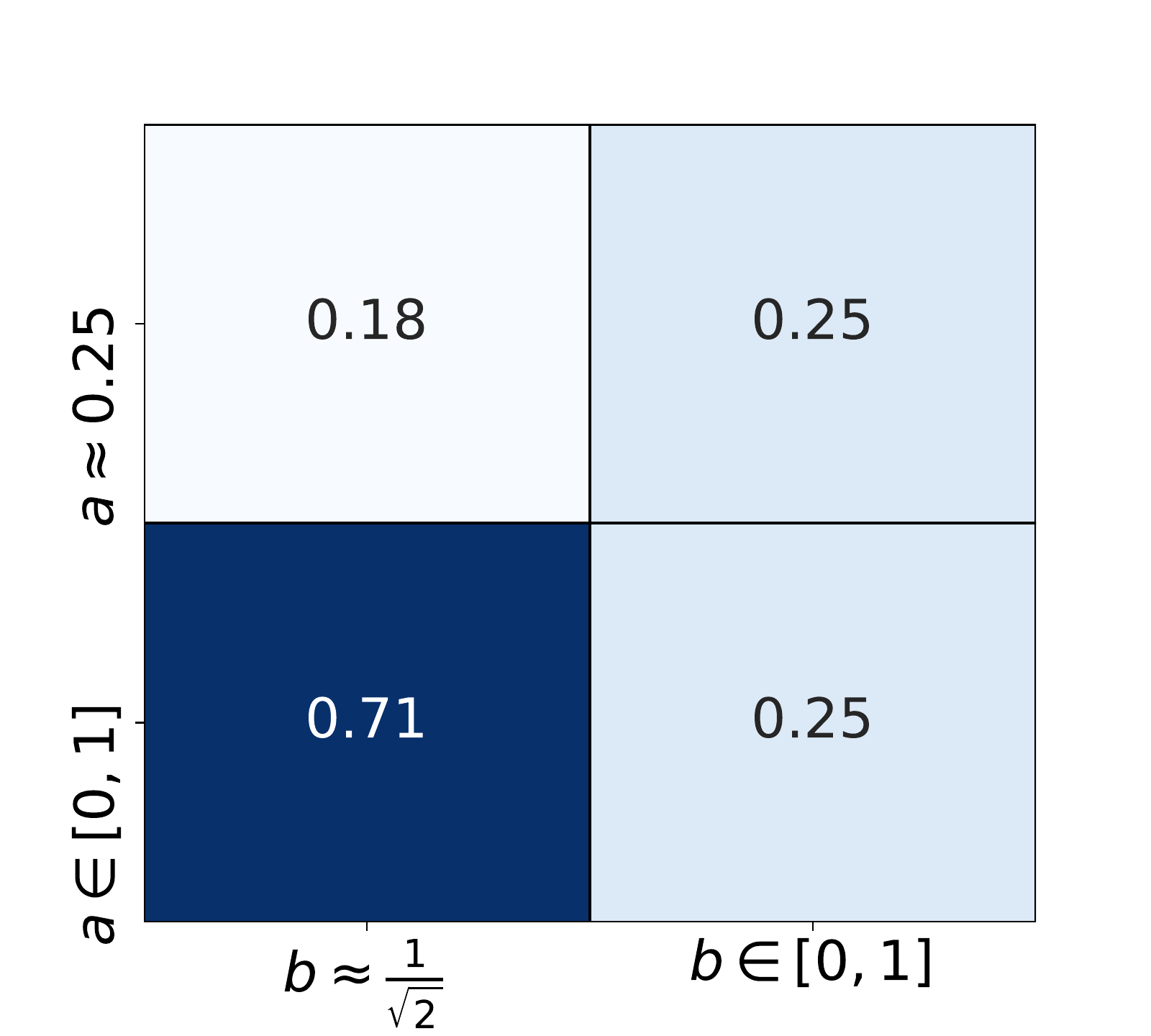}
 }
 \subfloat[{Success rate for trained circuit for classification of data}]{
   \includegraphics[width=0.45\linewidth]{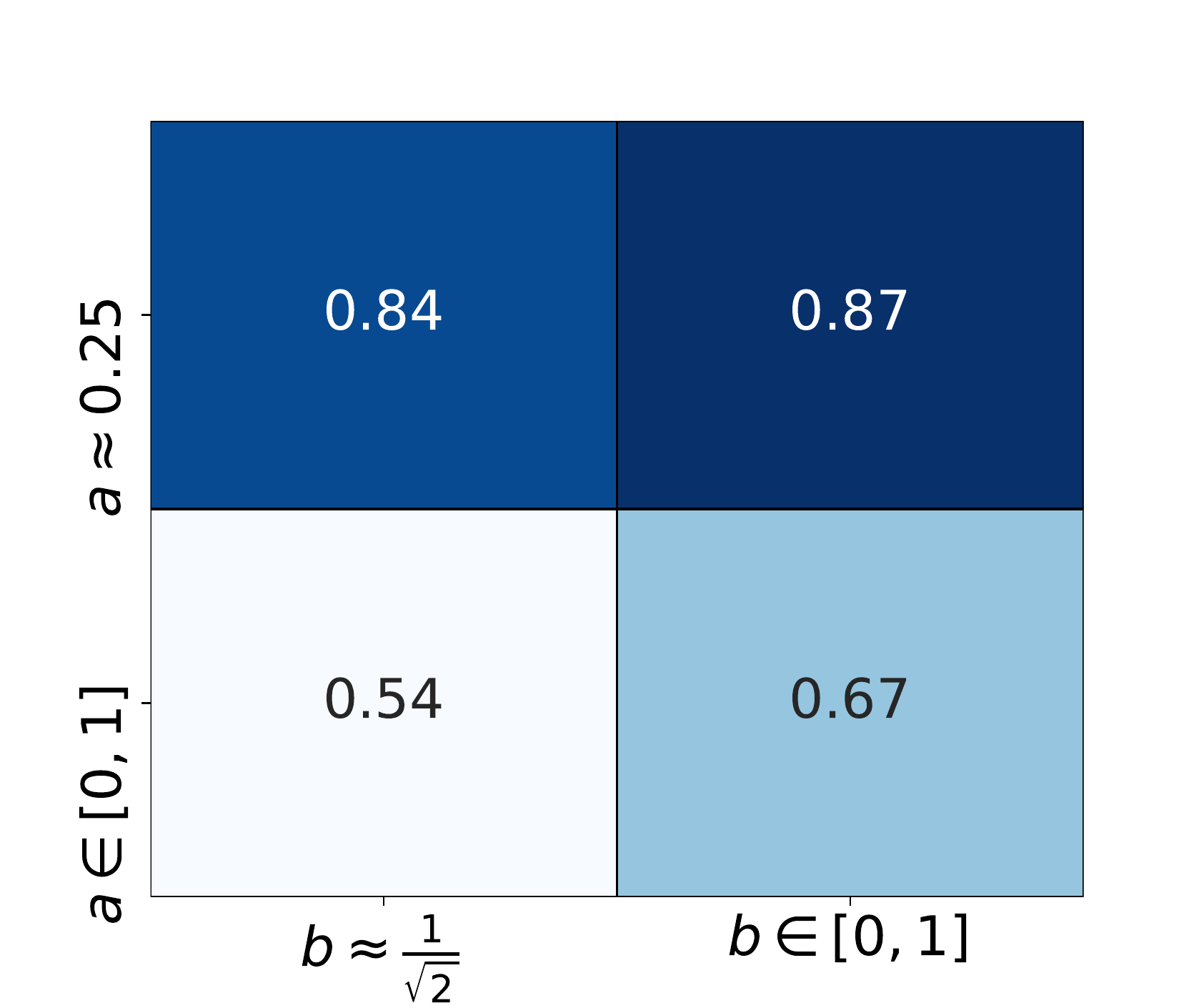}
 }
 \caption{Classification of data sampled from different non-orthogonal quantum probability distributions. Here we show that learned quantum circuits are capable of generalizing quantum data which is sampled from a variety of different join probability distributions for multiple inputs. The classification is done in an unambiguous manner (with error rate $<0.01$), and we find an inverse correlation between
 the success rate and the averaged fidelity between the input quantum states. The values $0.25$ or $\frac{1}{\sqrt{2}}$ respectively on the axis represents a normal
 distribution centred at corresponding value, with standard deviation $0.05$. The value $[0,1]$ represents a uniform distribution.
 The data is obtained as an average over $50$ repetitions.}
 \label{fig:ud_distributions}
\end{figure*}

\paragraph{Distinguishing data from different probability distributions.---}
Here we shown that our quantum circuit has the power
 to unambiguously classify data which was not seen during the optimisation process. We attempt to show this by allowing both parameters $a$ and $b$ to be drawn from
 some probability distribution. That is, we use training data which is sampled from some distribution, and test it performance with data sampled from the same distribution
 for both families of states. We show that the trained circuit can successfully classify data with $a$ and $b$ drawn from the normal distribution, the uniform distribution,
 or a mixture of two. The resulting success rate was inversely correlated with the averaged fidelity between the quantum states.
 The data distribution is shown in Fig.~\ref{fig:ud_distributions}.

%

\section{Learning convergence from ensemble measurements}
\label{sec:use_measurement}

Here we simulate the process that a classical-quantum hybrid scheme would implement utilizing a quantum device and analyse its performance.
These numerical simulations can in principle be validated in a physical experiment, where the measurement outcomes are used to infer the different probabilities for the cost function.
To have good estimation of the probability, and hence the  cost function, one has
to make repeated measurements to train the model, and we note that in particular better methods to evaluate the analytical gradient are available on a shallow quantum device~\cite{mitarai2018quantum}.
We first give a brief discussion of the an estimated number of repeated measurements which are required to approximate the gradient, which is oriented on~\cite{farhi2018classification}[Section 3].
Since the gradients are calculated using the forward difference formula:
\begin{equation}
  \frac{df}{dx}(x) = \frac{f(x+\varepsilon) - f(x)}{\varepsilon} +
  O(\varepsilon^2)
\end{equation}
The error in the calculation of $f$ must be at most of the order of
$O(\varepsilon^2)$, in order to prevent dominating the total error. To
achieve this ideally with an $99\%$ probability, one requires the number of repeated
measurement to be of the order $\frac{1}{(\varepsilon^2)^2}=\frac{1}{\varepsilon^4}$.\footnote{
This assumes that the cost function follows a normal distribution with variance of the order $\frac{1}{\sqrt{N}}$, where $N$ is the number of measurements made in reach run in order to calculate the cost function.
}
For example, when $\varepsilon = 10^{-3}$, the ideal number of repetitions is given by $10^{12}$.

In practice, we do not use $\frac{1}{\varepsilon^4}$ measurements, since
Adam is designed with the stochasticity of cost function taken into account. To give an estimate of the number
of repeated measurements which are required for convergence of the optimisation process,
we perform two numerical experiments. We first look at the case when the number of
repeated measurements is $\geq 10^3$ and $\varepsilon=10^{-2}$. We find that
$10^5$ repeated measurements for each iteration are a robust configuration
for successful convergence. Second, we vary the learning rate and increase the maximal number of iterations for Adam, setting $\varepsilon=10^{-2}$
and taking only $100$ repeated measurements. We observe that optimisation is successful with the large number of iteration. In both experiments, the penalties were set to $\alpha_{inc}=5$ and $\alpha_{err}=40$.

\begin{figure*}
\begin{center}
\includegraphics[width=0.5\linewidth]{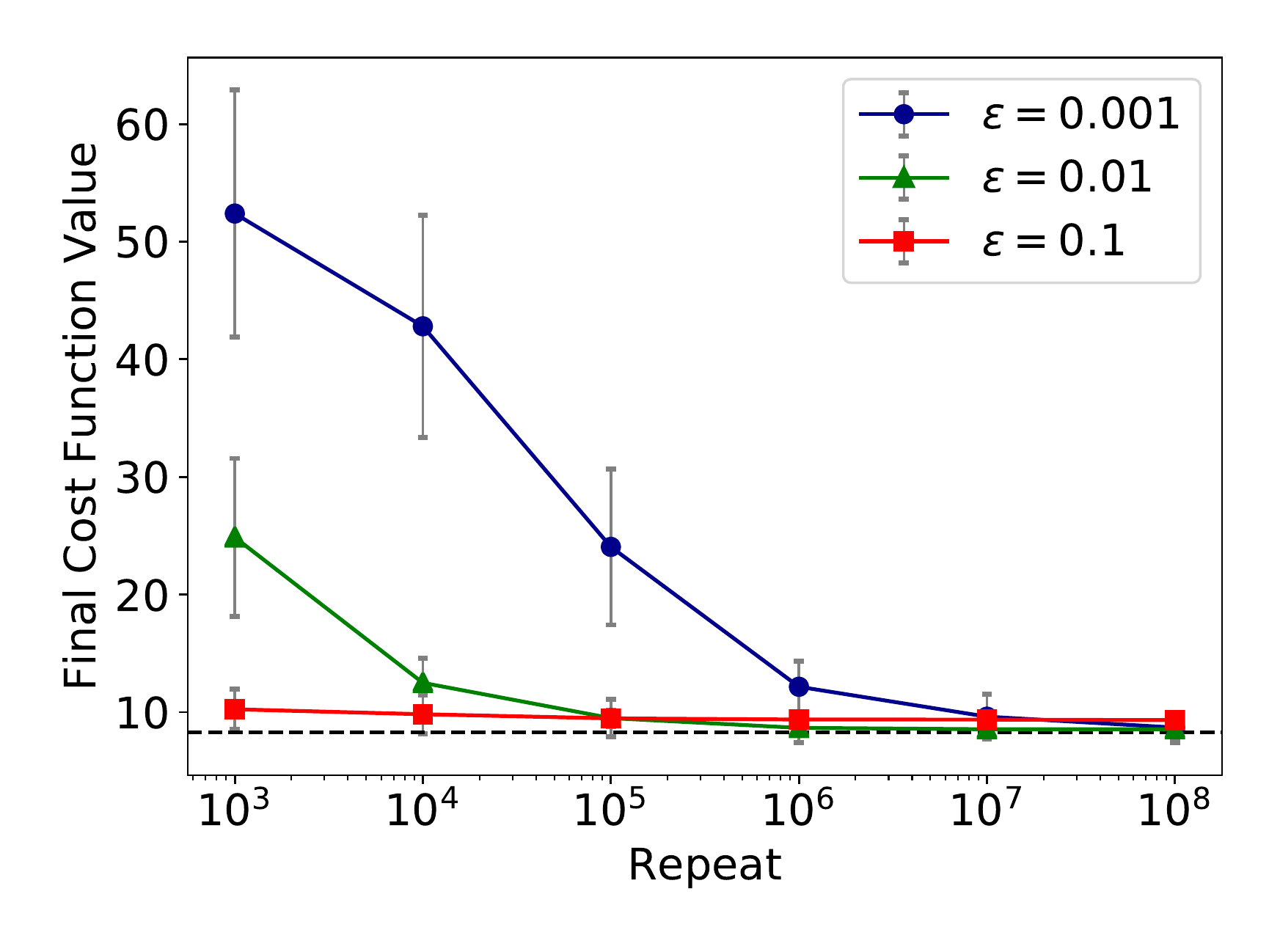}
\caption{The cost function after 5000 iterations. The result obtained using exact probabilities is shown by the horizontal dashed line.
  For a smaller step size for gradient ($\varepsilon$), we find that more repetitions are required to give a consistent result.
  However, a combination of $\varepsilon=10^{-2}$ and $10^5$ repetitions gives a result which well approximates
  the result obtained using exact probabilities.
  Here \textit{repeat} is the number of repeated measurements that are made each time to
  calculate the cost function. The cost function values are averaged over
  $50$ repeated runs of the training process, and the bars indicate the standard deviations.
}
\label{fig:repeat_m}
\end{center}
\end{figure*}

\paragraph{Large number of repetitions.} Our results show that for a fixed
maximum number of iterations ($5000$) for Adam, a combination of $\varepsilon=10^{-2}$ and
$10^5$ repeated measurements gives robust results, i.e., the final cost
function is close to the value obtained with the exact probability (with an
error within $3\%$) and is stable (with a relative standard deviation of
$13\%$). A more detailed description of the trade-off between repeated
measurements and the stability of the cost function is shown in
Fig.~\ref{fig:repeat_m}.

\begin{figure*}
\begin{center}
  \subfloat[{Lowering learning rate to counter the effect of insufficient 
  sampling. Both the value and the standard deivation of cost function at the 20000'th iteration were brounght down by lowered learning rate. Here the number of repeated measurements is only $100$, much 
  smaller than $\frac{1}{\varepsilon^4}=10^8$.}]{
    \includegraphics[width=0.5\linewidth]{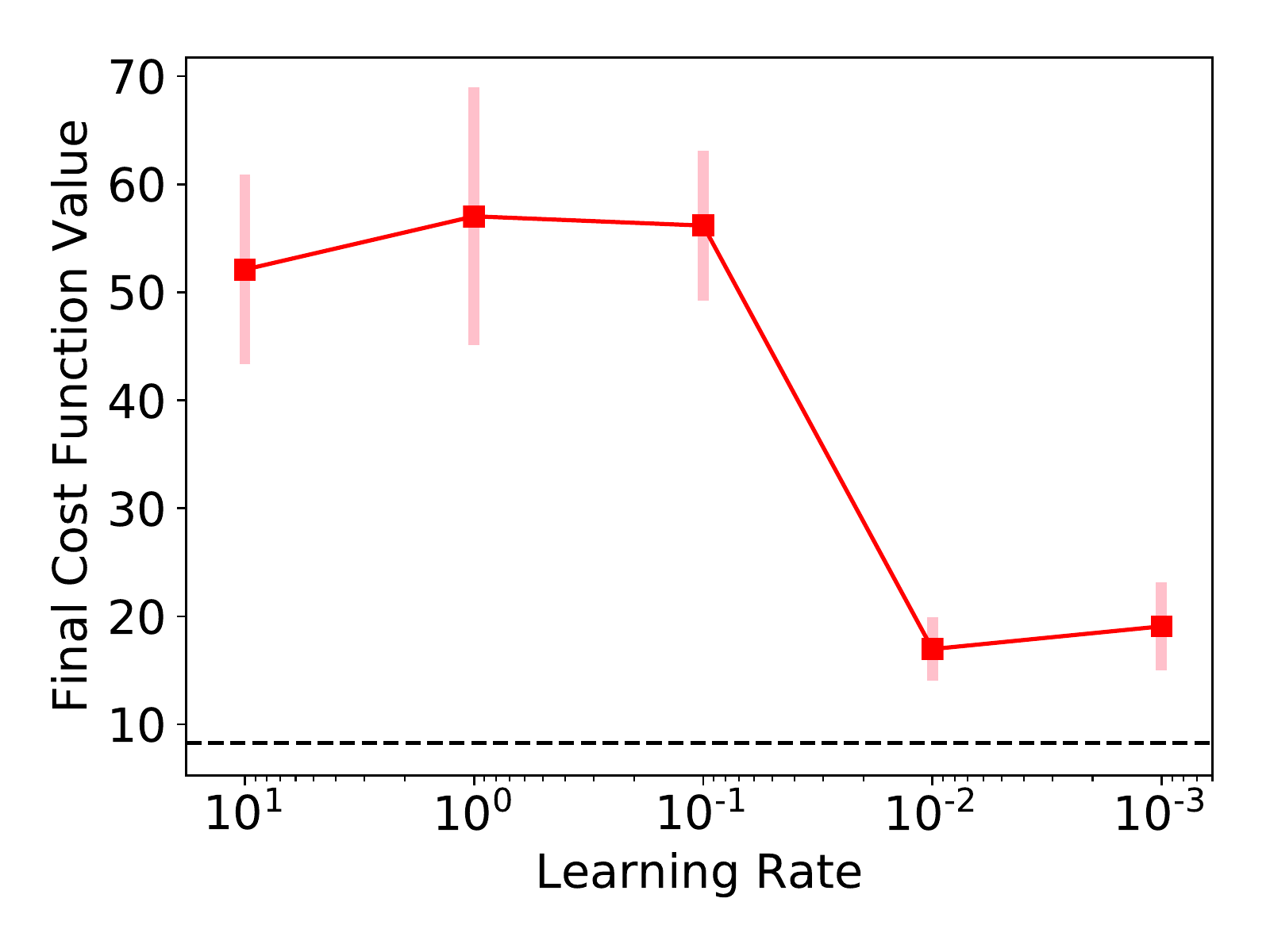}
  }
  \subfloat[{The noisy cost function $J_1$ estimated by $100$ repeated measurements slowly moved to its optimal value 
  when the learning rate is set as $0.001$. The horizontal dashed line showed the minimal value $8.3$ for cost function.
  The inset illustrate that trained circuit could discrminate the two quantum states. Although the error rate in the inset is not $0$, we believe it could be achieved by further tuning the penalities, which was set as ($\alpha_{err}=40$, $\alpha_{inc}=5$) for this training.}]{
  	\includegraphics[width=0.5\linewidth]{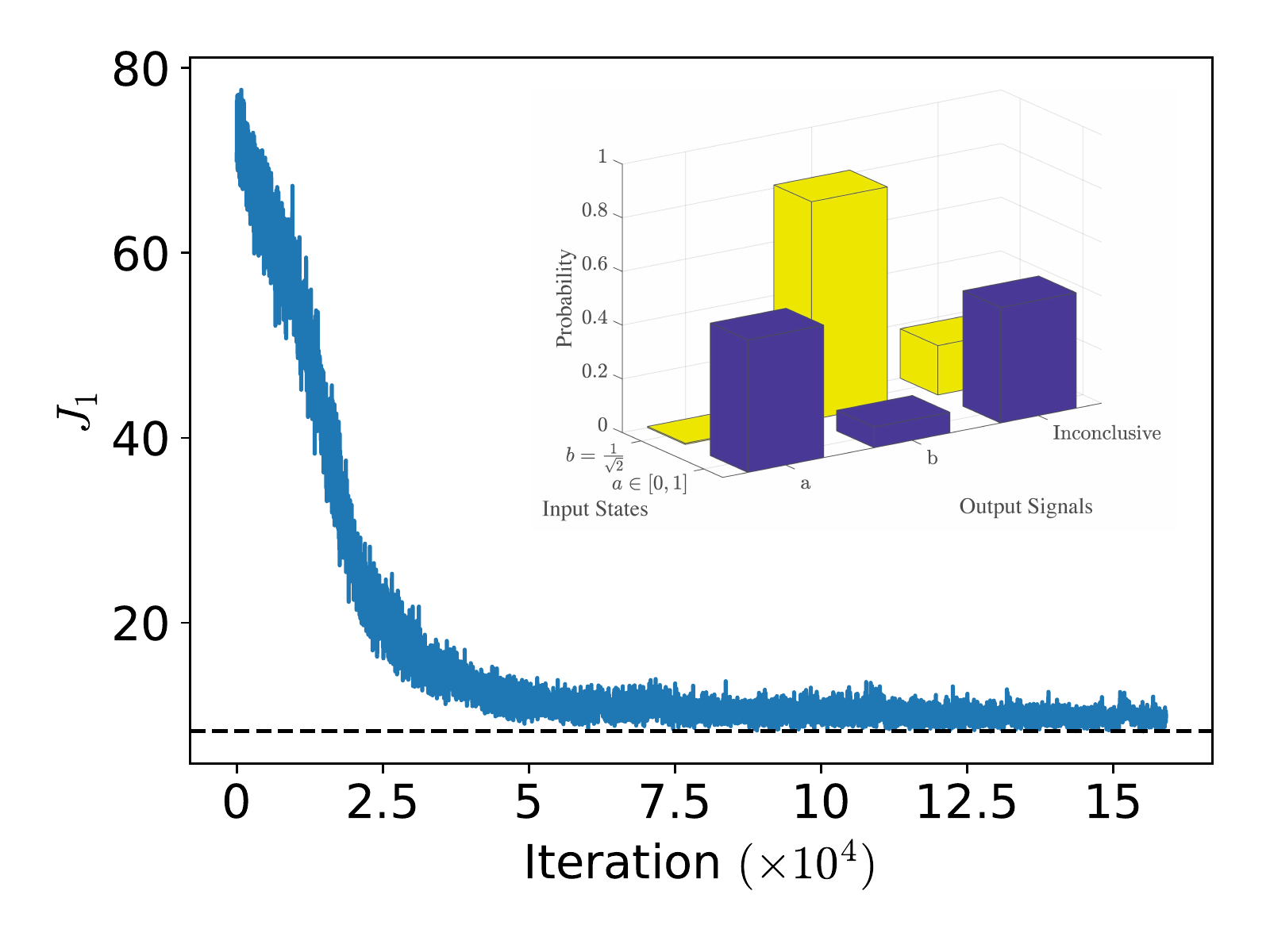}
  }
    \caption{Small learning rates with high number of iterations.}
\label{fig:repeat_m_small_geps}
\end{center}
\end{figure*}

\paragraph{Small learning rates and high number of iterations.}
Our numerical experiments further showed that in the case of small repetitions, lowered learning rate could effectively counter the noisy brought by the insufficient sampling.
Although in this case, the optimisation required a longer iteration to finish.
For example, with only 100 repeated measurements, the variance of cost function $J_1$ after 20000 iterations decreased as we lowered the learning rate (Fig.\ref{fig:repeat_m_small_geps}(a)). We could visually observe the optimisation process where the cost function $J_1$ slowly approached the optimal value in Fig~\ref{fig:repeat_m_small_geps}(b). Here, gradient step were taken as $\varepsilon=10^{-2}$.

\section{Conclusions}

We have developed a universal quantum circuit learning approach for discrimination and classification of quantum data. In particular, we have designed a
theoretically motivated cost function and then used the stochastic optimisation algorithm Adam in a quantum-classical hybrid scheme to train a circuit to perform quantum state discrimination. The training was performed over a prior specified range of input states, however, without training the circuit on the whole range. This training process generalised well for the discrimination task on new data, i.e., states from the parameter range which have not been seen during the training process. This in particular distinguishes our work from previous results on quantum circuit learning, in particular very recent study in  e.g.~\cite{Fanizza2018Optimal}, who only optimise circuits for specific inputs.  Note that prior work hence does not consider the generalisation ability and hence does not treat the actual learning problem which is aiming at optimisation as well as generalisation.\\
We observed a trade-off between error and inconclusiveness rates when we
penalised them differently in the cost function. Although this experiment was done on a
simulated quantum computers, i.e., classical hardware, where exact measurement
probabilities are available, we showed that this
optimisation could be experimentally performed with a repeated number of measurements of the wavefunction. Finally we note that recent quantum methods for estimating the analytical gradient via variations in the unitaries~\cite{mitarai2018quantum} can be directly applied for training our circuits and therefore one can perform the optimisation efficiently on near term quantum devices.

The discriminative quantum neural networks of the forms that were trained here could be potentially  used as quantum shallow circuits for verifying or certifying other shallow or deep quantum circuits within machine learning or quantum simulations applications. They could also be used to verify the output of other generative models, such as Restricted Boltzmann machines or GANS ~\cite{goodfellow2014generative}. Small-scale discriminative quantum circuit learning could be used for constructing non-trivial (e.g., POVM-based) receivers in quantum meteorology \cite{giovannetti2011advances}, sensing \cite{degen2017quantum}, and imaging  \cite{chen2012optical}.

\section{Acknowledgements}

We want to thank Raban Iten, Oliver Reardon-Smith and Roger Colbeck for valuable insights in parametrising the general measurement circuits and Jarrod McClean for feedback on the manuscript. This project acknowledges the use of the ESPRC funded Tier 2 facility JADE, the use of the UCL Legion High Performance Computing Facility (Legion@UCL), and associated support services, in the completion. L.W. is supported by the Royal Society. S.S. is supported by the Royal Society, EPSRC, the National Natural Science Foundation of China, and the grant ARO-MURI W911NF-17-1-0304 (US DOD, UK MOD and UK EPSRC under the Multidisciplinary University Research Initiative). This work has been carried out while L.W. and S.S. participated in the workshop of \textit{Measurement and control of quantum systems} at the Institut Henri Poincare. The financial support is kindly acknowledged. 

\clearpage

%% file: sections/appendix.tex
\section{Quantum Circuits for POVMs}
\label{sec:m_circ_parametrisation}
This section describes the parametrisation of the circuit \ref{eq:general-measurement-circuit}.

\subsection{Cosine-Sine Decomposition}
Here we mention the Cosine-Sine Decomposition of unitary matrix, which will be
frequently used in the following. For every unitary matrix $U\in \C^{2^n\times2^n}$, it can be
decomposed as:
\begin{equation}
  U_n = \begin{pmatrix}
    A_0 & 0 \\ 0 & A_1
  \end{pmatrix}
  \begin{pmatrix}
    C & -S \\ S & C
  \end{pmatrix}
  \begin{pmatrix}
    B_0 & 0 \\ 0 & B_1
  \end{pmatrix}
\end{equation}
where $A_0,A_1,B_0,B_1$ are unitary matrices of size ${2^{n-1}\times2^{n-1}}$,
$C$ and $S$ are real diagonal matrices of size ${2^{n-1}\times2^{n-1}}$
satisfying $C^2+S^2=\id$. It is written in the following circuit equivalence:
\begin{equation}
  \Qcircuit @C=.3em @R=0em @!R{
                & \qw  & \multigate{1}{U_n} & \qw 
                & \push{\rule{1em}{0em}} &
    \qw &
    \sgate{}{1} & \sgate{R_y}{1} & \sgate{}{1}
                &\qw
    \\
    \lstick{\scriptstyle n-1} & {\backslash}\qw & \ghost{U_n} & \qw
                 & \push{\rule{.2em}{0em}=\rule{.2em}{0em}} &
    {\backslash}\qw &
    \gate{U_{n-1}} & \gate{} & \gate{U_{n-1}}
                   &\qw
  }
\end{equation}
Where a box \framebox[.6em]{} represents the control part of a uniformly controlled
gate, see section IV of \cite{Iten2015} for details. In circuit \ref{eq:general-measurement-circuit}, the first qubit is initiated to be $\ket{0}$, so we have
\begin{equation}
  \Qcircuit @C=.5em @R=0em @!R{
    \lstick{\ket{0}}   & \qw  & \multigate{1}{U_n} & \qw 
                & \push{\rule{2em}{0em}} &
    \qw         & \sgate{R_y}{1} & \sgate{}{1}
                &\qw
    \\
    \lstick{\scriptstyle n-1} & {\backslash}\qw & \ghost{U_n} & \qw
                 & \push{\rule{.1em}{0em}=\rule{.3em}{0em}} &
    \gate{U_{n-1}} & \gate{} & \gate{U_{n-1}}
                   &\qw
  }
\end{equation}

\subsection{Decomposition of circuit \ref{eq:general-measurement-circuit}}
For a general measurement giving at most 4 measurement outcomes, we have the
following circuit representation:
\begin{equation} \label{eq:4POVMs-org}
    \Qcircuit @C=1em @R=.7em {
      \lstick{\ket{0}}    & \multigate{3}{V} & \measure{M_1} & \cw \\
      \lstick{\ket{0}}    & \ghost{V}        & \measure{M_2} & \cw \\
      \lstick{\ket{\phi}} & \ghost{V}        & \qw \\
      \lstick{\ket{\psi}} & \ghost{V}        & \qw
      }
\end{equation}
The first $V$ could be decomposed using the circuit equivalence on page 5 of
\cite{Iten2016} into:
\[ \Qcircuit @C=1em @R=.7em {
    \lstick{\ket{0}}      & \multigate{3}{R} & \sgate{}{1}       & \measure{M_1} & \cw \\
    \lstick{\ket{0}}      & \ghost{R}        & \multigate{2}{V'} & \measure{M_2} & \cw \\
      \lstick{\ket{\phi}} & \ghost{R}        & \ghost{V'}        & \qw \\
      \lstick{\ket{\psi}} & \ghost{R}        & \ghost{V'}        & \qw
} \]
where the $R$ gate does not act on the second qubit. Applying Cosine-Sine
Decomposition gives

\[
  \Qcircuit @C=1em @R=0.7em {
    \lstick{\ket{0}} & \qw  & \sgate{R_y}{2} &\sgate{}{2}
                     & \sgate{}{1}
                     & \measure{M_1} & \cw
    \\
    \lstick{\ket{0}} & \qw & \qw  & \qw
                     & \multigate{2}{V'}
                     & \measure{M_2} & \cw
    \\
    \lstick{\ket{\phi}} & \multigate{1}{U} & \sgate{}{1} & \multigate{1}{U}
                        & \ghost{V'}
                        & \qw
    \\
    \lstick{\ket{\psi}} & \ghost{U} & \gate{} & \ghost{U}
                        & \ghost{V'}
                        & \qw
  }
\]

The uniform controlled $V'$ and $U$ can be merged and put after measurement of
$M_1$ as:

\[
  \Qcircuit @C=1em @R=0.7em {
    \lstick{\ket{0}} & \qw  & \sgate{R_y}{2} 
                     & \measure{M_1}\cwx[1]
    \\
    \lstick{\ket{0}} & \qw & \qw  
                     & \multigate{2}{V''} &\measure{M_2} & \cw
    \\
    \lstick{\ket{\phi}} & \multigate{1}{U} & \sgate{}{1} 
                        & \ghost{V''} &\qw 
    \\
    \lstick{\ket{\psi}} & \ghost{U} & \gate{} 
                        & \ghost{V''} &\qw 
  }
\]

The first line of the circuit could be merged with the second line as follows:
\newcommand{\ucc}[1]{*+<.6em>{#1} \POS ="i","i"+UR;"i"+UL **\dir{-};"i"+DL 
**\dir{-};"i"+DR **\dir{-};"i"+UR **\dir{-},"i" \cw}
\begin{equation}\label{eq:4POVMs-1}
  \Qcircuit @C=0.5em @R=0.7em {
                  & &
                  & \cwx[1] &  \cctrl{1} & \cw  
                  & \ucc{} \cwx[1] 
                  & \cw
    \\
    \lstick{\ket{0}} & \qw & \sgate{R_y}{1}  
                     & \measure{M_1} & \targ & \qw 
                     & \multigate{2}{V''}
                     & \measure{M_2}
    \\
    \lstick{\ket{\phi}} & \multigate{1}{U} & \sgate{}{1} 
                        & \qw  & \qw  & \qw
                        & \ghost{V''}
                        & \qw
    \\
    \lstick{\ket{\psi}} & \ghost{U} & \gate{} 
                        & \qw  & \qw  & \qw
                        & \ghost{V''}
                        & \qw
  }
\end{equation}

And then we apply the Cosine-Sine Decomposition to $V''$, throwing away
the last gate on third and forth qubits, we obtain:

\begin{equation}\label{eq:4POVMs-2}
  \Qcircuit @C=0.3em @R=0.7em {
                  & &
                  & \cwx[1] &  \cctrl{1} & \cw  
                  & \ucc{} \cwx[2]  
                  & \ucc{} \cwx[1]  
                  & \cw
    \\
    \lstick{\ket{0}} & \qw & \sgate{R_y}{1}  
                     & \measure{M_1} & \targ & \qw \qw 
                     & \qw
                     & \gate{R_y}
                     & \measure{M_2}
    \\
    \lstick{\ket{\phi}} & \multigate{1}{U} & \sgate{}{1} 
                        & \qw  & \qw  & \qw
                        & \multigate{1}{U}
                        & \sgate{}{-1}
                        & \qw
    \\
    \lstick{\ket{\psi}} & \ghost{U} & \gate{} 
                        & \qw  & \qw  & \qw
                        & \ghost{U}
                        & \sgate{}{-1}
                        & \qw
  }
\end{equation}

The uniformly-controlled rotations and remaining two qubit unitary gates could be easily parametrised by CNOTs and single qubit rotations. For example, see \cite{Shende2006} and \cite{Shende2004a}.


\section{Adam algorithm for stochastic optimisation}
\label{sec:adam_intro}

Here we provide a brief introduction to the Adam algorithm for stochastic
optimisation. Adam is based on the gradient descent algorithm for function
optimisation.  The gradient descent algorithm starts with an initial guess of
minimal parameter $\theta_1$ and updates this parameter iteratively until a
minimal value is obtained by the following rule:
\begin{equation}
  \theta_{t} = \theta_{t-1} - \alpha \grad_{\theta} J(\theta_{t-1})
\end{equation}
Here $J$ is the cost function, which could be stochastic. $\alpha$ is called
the learning rate and its value requires empirically tuning. We used two
improvements on gradient descent to optimise our circuit.

\paragraph{Stochastic calculation of gradient.} In practice, the cost function
$J$ often has the following form:
\begin{equation}
  J(\theta) = \frac{1}{N} \sum_{i=1}^N J_i(\theta)
\end{equation}
Here each $J_i$ is usually associated with a single datum for optimisation.  The
calculation of gradient may be computationally expensive when $N$ is large, in
which case we calculated it in a stochastic manner. Specifically, at each
calculation of gradient, we sampled a mini-batch
$B=\{i'_1,i'_2,\cdots,i'_{N'}\}$, drawn uniformly from the training data, and
calculated an estimate of gradient in this mini-batch:
\begin{equation}
  g_t = \frac{1}{N'} \sum_{i'\in B} \grad J_{i'}(\theta)
\end{equation}
The $N'$ was held fixed throughout the training.

\paragraph{Adaptive moment estimation (Adam).} Adam improves on the gradient
descent by incorporating two pieces of information:
\begin{align}
  m_t &= \beta_1 m_{t-1} +  (1-\beta_1) g_t \\
  v_t &= \beta_2 v_{t-1} + (1-\beta_2) g_t^2
\end{align}
Here $g_t = \grad_{\theta}J(\theta_t)$, and $g_t^2=g_t \odot g_t$ is a vector of
element-wise squares of gradient. The first term $m_t$ is an exponential moving
average of the gradient controlled by parameter $\beta_1$, and the second term
$v_t$ is an exponential moving average of the $g_t^2$ controlled by parameter
$\beta_2$. These two values are combined in updating the parameter $\theta$ in
the following manner:
\begin{equation}
  \theta_{t} = \theta_{t-1} - \alpha \frac{m_t}{\sqrt{v_t} + \hat{\varepsilon}}
  \label{eq:adam_simple}
\end{equation}
Here $\hat{\varepsilon}$ is a small number to avoid division by zero when $v_t$ is
initialised to be $0$.

With $m_t$, the parameter update in Eq.~\ref{eq:adam_simple} will favor the direction
where the gradient points mostly to the same direction, while disfavor
direction where the gradient oscillates backwards and forwards. Intuitively,
$m_t$ makes the cost function $J$ ``accelerates" in the optimisation by
accumulating ``momentum" $g_t$. 

With $v_t$ in Eq.~\ref{eq:adam_simple}, a moving average of the magnitudes of
gradient in each direction is included, and the direction with smaller gradient
is amplified in the update. Intuitively, this amplifies the influence of rarely
seen features (which contribute to small gradients) on the training.

In practice, we initialised $m_t$ and $v_t$ to be a zero vector, which
made the moving averages biased towards zero. This problem was corrected by the
following adjusted updating rules (as suggested in the original
paper\cite{Kingma2014}):
\footnote{
  Note that we used here the formulation mentioned just before Section 2.1 of the paper
  \cite{Kingma2014}, which improves the efficiency of Algorithm~1 in that paper.
}:
\begin{algorithmic}
\While{$\theta_t$ not converged} 
  \State $t \gets t+1$
  \State $\alpha_t \gets \alpha \vdot \sqrt{1-\beta_2^t} / (1-\beta_1^t)$
  \State $g_t \gets \grad_{\theta} J(\theta_{t-1})$
  \State $m_t \gets \beta_1 m_{t-1} + (1-\beta_1) g_t$
  \State $v_t \gets \beta_2 v_{t-1} + (1-\beta_2) g_t^2$
  \State $\theta_t \gets \theta_{t-1} - \alpha_t \vdot m_t / (\sqrt{v_t}
  +\hat{\varepsilon})$
\EndWhile
\end{algorithmic}

Generally, Adam is very robust to the choice of parameters, and a good choice of
parameters suggested by its authors are: $\alpha=0.001$, $\beta_1 = 0.9$,
$\beta_2 = 0.999$, and $\hat{\varepsilon} = 10^{-8}$.

\section{Discussions on stochastic optimization strategies}
\label{sec:optimiser}
\begin{figure}[th]
	\centering
	\includegraphics[width=0.7\linewidth]{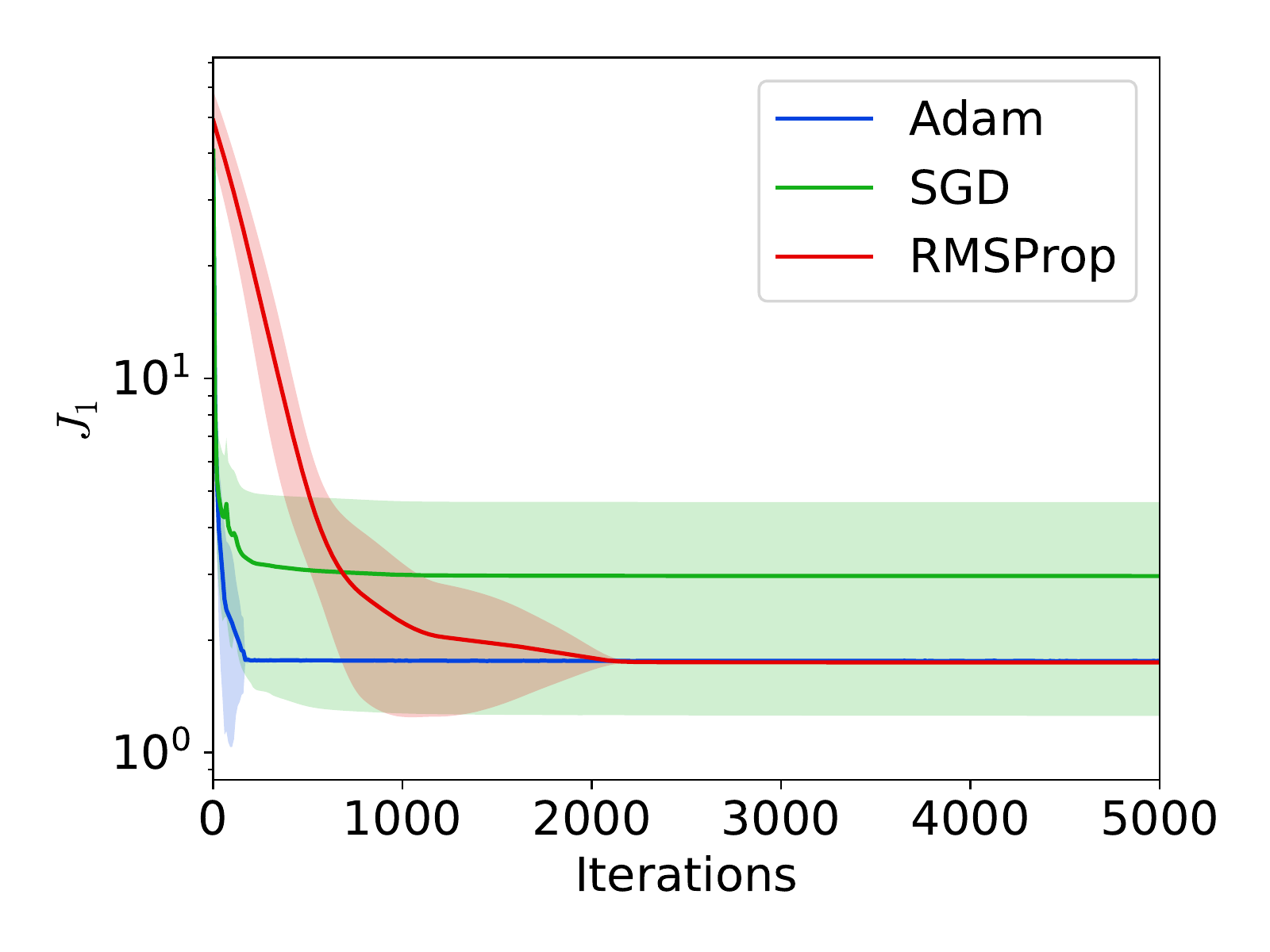}
	\caption{
		Performance curve of different optimisation algorithms, where SGD often stops at a plateau of higher cost functions. Here the shading indicates. The shading represents one standard deviation
		computed across 10 runs from random initial parameters.}
	\label{fig:diffopt}
\end{figure}

In our work we observed a poor performance for stochastic gradient descent (SGD). However, when replacing SGD with the Adam algorithm, we are able to recover a good solution, i.e., nearly optimal results. Recent result by \cite{mcclean2018barren} imply that classical-quantum hybrid methods might not perform well in practice based on a proof involving random unitaries.
However, we showed here that the optimisation procedure works well for the distributions of qubits studied if we replace the stochastic gradient descent with improved methods like Adam, or RMSProp~\cite{tieleman2012rmsprop}(c.f. Fig.\ref{fig:diffopt}). Since it is in general hard to determine the concrete reasons for failure of the optimisation process, based on the resulting performance of different optimisation algorithms, we can hence only hypothesise about origins of the observed behaviour.  
One explanation is the usage of non-optimal training parameters. A solution to this would be to perform an optimisation on these parameters which we didn't include in this work.\\

Another explanation is the saddle-point hypothesis. Since methods like Adam and RMSProp have been shown to perform particularly well in high-dimensional landscapes we suggest that the widely believed proliferation of saddle-points hypothesis~\cite{martens2010deep,dauphin2014identifying,arjovsky2015saddle,sagun2016eigenvalues,wossnig2017classical} might also apply to quantum circuit training. 
However, in many practical applications stochastic gradient descent has been shown to perform well and been able to escape saddle-points due to its stochastic component, while only certain cases restrict its usage.\\

In comparison with our structure, in Ref. \cite{mcclean2018barren} the authors also assumes a certain type of unitary circuit, i.e., i.i.d. randomly sampled circuits which form a random unitary matrix. 
Referring to Spielman and Tengs monograph on the smoothed analysis~\cite{spielman2004smoothed},
we conjecture that the special case of randomly sampled matrices does not apply in our highly structured problem, and support this hypothesis with the observation that up to the number of qubits we obtain a stable optimisation process. However, we leave open whether this holds for larger amount of qubits.

